\documentclass[12pt,a4paper,twoside]{article}
\usepackage[a4paper,tmargin=3truecm,bmargin=3truecm,rmargin=2.2truecm,lmargin=2.2truecm]{geometry}
\pdfoutput=1
\usepackage[T1]{fontenc}
\usepackage[utf8]{inputenc}
\usepackage{amssymb, amsmath}
\usepackage[all]{xy}
\usepackage{graphicx} 
\usepackage{stmaryrd}
\usepackage[amsmath, hyperref, thmmarks]{ntheorem}
\usepackage[english]{babel}
\usepackage[pdftex]{hyperref}

\let\tr\relax 

\newcommand\varnotation[1]{{\mathcal{#1}}}

\newcommand\algnotation[1]{{\mathbf{#1}}}

\newcommand\grnotation[1]{{\mathsf{#1}}}

\newcommand\evnotation[1]{{\mathnormal{#1}}}

\newcommand\modnotation[1]{{\boldsymbol{#1}}}

\newcommand\faiscnotation[1]{{\mathcal{#1}}}

\newcommand\varE{{\varnotation{E}}}
\newcommand\varF{{\varnotation{F}}}
\newcommand\varM{{\varnotation{M}}}
\newcommand\varN{{\varnotation{N}}}
\newcommand\varP{{\varnotation{P}}}
\newcommand\varQ{{\varnotation{Q}}}
\newcommand\varR{{\varnotation{R}}}

\newcommand\algA{{\algnotation{A}}}
\newcommand\algB{{\algnotation{B}}}
\newcommand\algC{{\algnotation{C}}}

\newcommand\algW{{\algnotation{W}}}

\newcommand\algzero{{\grnotation{0}}}
\newcommand\algun{{\grnotation{1}}}

\newcommand\evF{{\evnotation{F}}}

\newcommand\evV{{\evnotation{V}}}

\newcommand\modC{{\modnotation{C}}}

\newcommand\modE{{\modnotation{E}}}
\newcommand\modK{{\modnotation{K}}}
\newcommand\modM{{\modnotation{M}}}

\newcommand\faiscF{{\faiscnotation{F}}}

\newcommand\faiscH{{\faiscnotation{H}}}

\newcommand\bbbone{{ \mathchoice {1\mskip-4mu\mathrm{l} } {1\mskip-4mu\mathrm{l} }{1\mskip-4.5mu\mathrm{l} } {1\mskip-5mu\mathrm{l}} }}

\newcommand\gR{{\mathbb R}}

\newcommand\gF{{\mathbb F}}
\newcommand\gC{{\mathbb C}}

\newcommand\gN{{\mathbb N}}

\newcommand\caA{{\mathcal A}}

\newcommand\caF{{\mathcal F}}
\newcommand\caG{{\mathcal G}}

\newcommand\caI{{\mathcal I}}

\newcommand\caN{{\mathcal N}}

\newcommand\caP{{\mathcal P}}

\newcommand\caX{{\mathcal X}}
\newcommand\caY{{\mathcal Y}}
\newcommand\caZ{{\mathcal Z}}

\newcommand\ka{{\mathfrak a}}
\newcommand\kb{{\mathfrak b}}
\newcommand\kg{{\mathfrak g}}
\newcommand\kh{{\mathfrak h}}
\newcommand\ki{{\mathfrak i}}
\newcommand\kk{{\mathfrak k}}
\newcommand\kl{{\mathfrak l}}
\newcommand\km{{\mathfrak m}}
\newcommand\kn{{\mathfrak n}}
\newcommand\ks{{\mathfrak s}}
\newcommand\kz{{\mathfrak z}}

\newcommand\kS{{\mathfrak S}}

\newcommand\kU{{\mathfrak U}}

\newcommand\kX{{\mathfrak X}}
\newcommand\kY{{\mathfrak Y}}

\newcommand\ksl{{\mathfrak{sl}}}

\newcommand\ksu{{\mathfrak{su}}}

\newcommand{\omi}[1]{\buildrel { \buildrel{#1}\over{\vee} } \over .}

\newcommand\ensvide{{\varnothing}} 
\newcommand\exter{{\textstyle\bigwedge}} 
\newcommand\symes{{\mathchoice{\textstyle\mathsf{S}}{\textstyle\mathsf{S}}%
{\scriptstyle\mathsf{S}}{\scriptscriptstyle\mathsf{S}}}} 

\newcommand\Ad{{\text{\textup{Ad}}}} 

\newcommand\der{{\text{\textup{Der}}}}

\newcommand\equ{{\text{\textup{equ}}}}

\newcommand\Int{{\text{\textup{Int}}}}

\newcommand\lie{{\text{\textup{Lie}}}}
\newcommand\loc{{\text{\textup{loc}}}}

\newcommand\Out{{\text{\textup{Out}}}}

\newcommand\sign{{\text{\textup{sign}}}}

\newcommand{\grast}{\bullet}

\newcommand\cdotaction{\mathord{\cdot}}

\DeclareMathOperator{\Aut}{Aut} 
\DeclareMathOperator{\End}{End} 
\DeclareMathOperator{\Hom}{\mathsf{Hom}} 
\DeclareMathOperator{\Ker}{Ker} 
\renewcommand{\ker}{\Ker} 
\DeclareMathOperator{\tr}{Tr} 

\newcommand{\super}[2]{\setbox1=\hbox{$#1$} \dimen2=\ht1 \dimen3=\dp1
                \setbox2=\hbox{$#2$}
                \dimen1=\wd1 \advance\dimen1 by -\wd2 \divide\dimen1 by 2
                \advance\dimen1 by \wd2 
                \setbox3=\hbox to \wd1{\hss \box1 \kern -\dimen1 \box2 \hss}
                \ht3=\dimen2 \dp3=\dimen3
                \box3
               }
               
\newcommand{\Super}[2]{
               \setbox2=\hbox{$#1$}
               \setbox3=\hbox{$#2$} \dimen2=\ht3 \dimen3=\dp3 
               \dimen1=\dp2 \advance\dimen1 by -11.4pt 
               \advance\dimen1 by \dimen2
               \setbox4=\hbox to \wd3{\hss \box2 \hss}
               \setbox1=\hbox to \wd3{\vbox{\box4\vglue \dimen1\box3} \hss}
               \dp1=\dimen3 \ht1=\dimen2
               \box1
              }
              
\newcommand{\barre}[1]{\setbox1=\hbox{$#1$} \dimen2=\ht1 \dimen3=\dp1 \dimen4=\wd1
                \setbox2=\hbox{\sl /}
                \dimen1=\wd1 \advance\dimen1 by -\wd2 \divide\dimen1 by 2
                \advance\dimen1 by \wd2 \advance\dimen1 by 0.4pt
                \setbox3=\hbox to \wd1{\hss \box1 \kern -\dimen1 \box2 \hss}
                \ht3=\dimen2 \dp3=\dimen3 \wd3=\dimen4
                \box3
               }

\newcommand\adrep{{\text{\textup{ad}}}} 

\theoremnumbering{arabic}
\theoremstyle{break}
\theoremsymbol{}
\theorembodyfont{\slshape}
\theoremheaderfont{\normalfont\bfseries}
\theoremseparator{}
\newtheorem{Theorem}{Theorem}[section]
\newtheorem{theorem}[Theorem]{Theorem}
\newtheorem{Proposition}[Theorem]{Proposition}
\newtheorem{proposition}[Theorem]{Proposition}

\newtheorem{lemma}[Theorem]{Lemma}

\theorembodyfont{\upshape}
\theoremsymbol{\ensuremath{\blacklozenge}}
\newtheorem{Example}[Theorem]{Example}
\newtheorem{example}[Theorem]{Example}
\newtheorem{Remark}[Theorem]{Remark}
\newtheorem{remark}[Theorem]{Remark}

\newtheorem{definition}[Theorem]{Definition}

\theoremstyle{nonumberplain}
\theoremheaderfont{\scshape}
\theorembodyfont{\normalfont}
\theoremsymbol{\ensuremath{\blacksquare}}

\newtheorem{proof}{Proof}
\qedsymbol{\ensuremath{_\blacksquare}}
\theoremclass{LaTeX}

\usepackage{stylearticle}

\newcommand\dd{\text{\textup{d}}}
\newcommand\hd{\widehat{\dd}}
\newcommand\hR{\widehat{R}}
\newcommand{\fibre}[4][]{\ensuremath{\xymatrix@1@C=14pt{{#2} \ar[r] & {#3} \ar[r]^-{#1} & {#4}}}}

\begin{document}

\title{Noncommutative generalization of $SU(n)$-principal fiber bundles: a review}
\author{T. Masson}
\date{}
\address{Laboratoire de Physique Th\'eorique (UMR 8627)\\
B\^at 210, Universit\'e Paris-Sud Orsay\\
F-91405 Orsay Cedex}
\ead{thierry.masson@u-psud.fr}


\begin{abstract}
This is an extended version of a communication made at the international conference ``Noncommutative Geometry and Physics'' held at Orsay in april 2007. In this proceeding, we make a review of some noncommutative constructions connected to the ordinary fiber bundle theory. The noncommutative algebra is the endomorphism algebra of a $SU(n)$-vector bundle, and its differential calculus is based on its Lie algebra of derivations. It is shown that this noncommutative geometry contains some of the most important constructions introduced and used in the theory of connections on vector bundles, in particular, what is needed to introduce gauge models in physics, and it also contains naturally the essential aspects of the Higgs fields and its associated mechanics of mass generation. It permits one also to extend some previous constructions, as for instance symmetric reduction of (here noncommutative) connections. From a mathematical point of view, these geometrico-algebraic considerations highlight some new point on view, in particular we introduce a new construction of the Chern characteristic classes.
\end{abstract}

\vfill
\begin{flushleft}
LPT-Orsay/07-57
\end{flushleft}
\newpage


\section{Introduction}

The geometry of fiber bundles is now widely used in the physical literature, mainly through the concept of connections, which are interpreted as gauge fields in particle physics. It is worth to recall why the structure of these gauge theories leads to this mathematical identification. The main points which connect these two concepts are the common expression for gauge transformations and the field strength of the gauge fields recognized as the curvature of the connection.

Since the introduction of the Higgs mechanics, some attempts have been made to understand its geometrical origin in a same satisfactory and elegant way as the gauge fields. The reduction of some higher dimensional gauge field theories to some more ``conventional'' dimensions has been proposed to reproduce the Higgs part of some models. 

Nevertheless, one of the more convincing constructions from which Higgs fields emerged naturally and without the need to perform some dimensional reduction of some extra \textsl{ad-hoc} afterward arbitrary distortion of the model, was firstly exposed in \cite{DuViKernMado:90b}, and highly popularized in subsequent work by A.~Connes in its noncommutative standard model (see \cite{ChamConn:07} for a review of the recent developments in this direction). What the pioneer work by Dubois-Violette, Kerner and Madore revealed is that the Higgs fields can be identified with the purely noncommutative part of a noncommutative connection on an noncommutative algebra ``containing'' an ordinary smooth algebra of functions over a manifold and a purely noncommutative algebra.

The algebra used there is the tensor product $C^\infty(\varM) \otimes M_n$ of smooth functions on some manifold $\varM$ and the matrix algebra of size $n$. This ``trivial'' product does not reveal the richness of this approach when some more intricate algebra is involved. In this review, we consider the algebra of endomorphisms of a $SU(n)$-vector bundle. This algebra reduces to the previous situation for a trivial vector bundle. This non triviality gives rise to some elegant and powerful constructions we exposed in a series of previous papers, and to some results nowhere published before.

The first part deals with some reviews of the ordinary geometry of fiber bundles and connections. We think this is useful to fix notations, but also to highlight what the noncommutative differential geometry defined in the following extends from these constructions.

We then define the general settings of our noncommutative geometry, which is based on derivations. The notion of noncommutative connections is exposed, and some important examples are then given to better understand the general situation.

The algebra we are interested in is then introduced as the algebra of endomorphisms of a $SU(n)$-vector bundle. We show how it is related to ordinary geometry, and how ordinary connections plays an essential role to study its noncommutative geometry.

The noncommutative connections on this algebra are then studied, and here we recall why the purely noncommutative part can be identified with Higgs fields.

Then it is shown that this algebra is indeed related, through the algebraic notion of Cartan operations on a bigger algebra, to the geometry of the $SU(n)$-principal fiber bundle underlying the geometry of the $SU(n)$-vector bundle.

Some considerations about the cohomology behind the endomorphism algebra are then exposed, in particular a new construction of the Chern classes of the $SU(n)$-vector bundle which are obtained from a short exact sequence of Lie algebras of derivations.

The last section is concerned with the symmetric reduction of noncommutative connections, which generalizes a lot of previous works about symmetric reduction of ordinary connections.

\section{A brief review of ordinary fiber bundle theory}
\label{Abriefreviewofordinaryfiberbundletheory}

The noncommutative geometry we will consider in the following contains, and relies in an essential way to the ordinary differential geometry of the $SU(n)$-fiber bundles theory. This section is devoted to some aspects of this differential geometry. Its aim is to fix notations but also to present some constructions which will be generalized or completed by the noncommutative geometry introduced later on.

\subsection{Principal and associated fiber bundles}

Let $\varM$ be a smooth manifold and $G$ a Lie group. Denote by $\fibre[\pi]{G}{\varP}{\varM}$ a (locally trivial) principal fiber bundle for the right action of $G$ on $\varP$, denoted by $p \mapsto p \cdotaction g = \widetilde{R}_g p$.

For any $p \in \varP$, one defines $V_p = \ker (T_p \pi: T_p \varP \rightarrow T_{\pi(p)}\varM)$, the vertical subspace of $T_p\varP$. For any $X \in \kg$, let
\begin{equation*}
X^v(p) = \left(\frac{d \hfill}{dt} p\cdotaction \exp(tX)\right)_{t=0}
\end{equation*}
Then $V_p = \{ X^v(p) / X \in \kg \}$ and one has $\widetilde{R}_{g \ast} V_p = V_{p\cdotaction g}$.
 
This defines vertical vector fields over $\varP$ and horizontal differential forms, which are differential forms on $\varP$ which vanish when one of its arguments is vertical.

Let $(U, \phi)$ be a local trivialisation of $\varP$ over an open subset $U \subset \varM$, which means that there exists a isomorphism $\phi : U \times G \xrightarrow{\simeq} \pi^{-1}(U)$ such that $\pi( \phi(x,h))=x$ and $\phi(x, hg) = \phi(x,h)\cdotaction g$ for any $x\in U$ and $g,h \in G$.

If $(U_i, \phi_i)$ and $(U_j, \phi_j)$ are two local trivialisations such that $U_i \cap U_j \neq \ensvide$, then there exists a differentiable map $g_{ij} : U_i\cap U_j \rightarrow G$ such that, if $\phi_i(x, h_i) = \phi_j(x, h_j)$ for $h_i, h_j \in G$, then $h_i = g_{ij}(x) h_j$ for any $x \in U_i \cap U_j$. The $g_{ij}$ are called the transition functions for the system $\{ (U_i, \phi_i) \}_{i}$ of local trivialisations. They satisfy $g_{ij}(x) = g_{ji}^{-1}(x)$ for any $x \in U_i \cap U_j$ and the cocycle condition $g_{ij}(x) g_{jk}(x) g_{ki}(x) = e$ for any $x \in U_i \cap U_j \cap U_k \neq \ensvide$.

Now, let $\varF$ be a manifold on which $G$ acts on the left: $\varphi \mapsto \ell_g \varphi$. On the manifold $\varP \times \varF$ we define the right action $(p,\varphi) \mapsto (p \cdotaction g, \ell_{g^{-1}} \varphi)$, and we denote by $\varE = (\varP\times \varF)/G$ the orbit space for this action. This is the associated fiber bundle to $\varP$ for the couple $(\varF, \ell)$. It is denoted by $\varE = \varP \times_\ell \varF$, and $[p,\varphi] \in \varE$ is the projection of $(p,\varphi)$ in the quotient $\varP \times \varF \rightarrow (\varP\times \varF)/G$. By construction, one has $[p \cdotaction g,\varphi] = [p, \ell_{g} \varphi]$.

A (smooth) section of $\varE$ is a (differentiable) map $s : \varM \rightarrow \varE$ such that $\pi \circ s(x) = x$ for any $x \in \varM$. We denote by $\Gamma(\varE)$ the space of differentiable sections of $\varE$.

The main point of this construction is the fact that one can show that $\Gamma(\varE)$ identifies with the space $\caF_G(\varP,\varF) = \{ \Phi : \varP \rightarrow \varF\ / \ \Phi(p\cdotaction g) = \ell_{g^{-1}} \Phi(p)\}$ of $G$-equivariant maps $\varP \rightarrow \varF$. This result will be useful later on.

Let $(U, \phi)$ be a local trivialisation of $\varP$ over $U$. Then the smooth map $s_U : U \rightarrow \pi^{-1}(U)$ given by $s_U(x) = \phi(x,e)$ is a local section of $\varP$.

Any section $s$ of $\varP$ is locally given by a local map $h : U \rightarrow G$ such that $s(x) = s_U(x)\cdotaction h(x) = \phi(x,h(x))$. We call $s_U$ a local gauge over $U$ for $\varP$, because it is used as a local reference in $\varP$ to decompose sections on $\varP$.

In the same way, any section $s$ of $\varE$ is locally given by a local map $\varphi : U \rightarrow \varF$ such that $s(x) = [s_U(x), \varphi(x)]$. This means that the local gauge $s_U$ can also be used to decompose sections of any associated fiber bundle.

Let $(U_i, \phi_i)$ and $(U_j, \phi_j)$ be two local trivialisations such that $U_i \cap U_j \neq \ensvide$. Then one has
\begin{equation*}
s_j(x) = \phi_j(x,e) = \phi_i(x, g_{ij}(x)) = \phi_i(x, e)\cdotaction g_{ij}(x) = s_i(x)\cdotaction g_{ij}(x)
\end{equation*}
so that on $\varP$, if $s(x) = s_i(x) \cdotaction h_i(x) = s_j(x) \cdotaction h_j(x)$, then 
\begin{equation*}
h_i(x) = g_{ij}(x) h_j(x)
\end{equation*}
On $\varE$, if $s(x) = [s_{i}(x), \varphi_i(x)] = [s_{j}(x), \varphi_j(x)]$ for $x \in U_i \cap U_j \neq \ensvide$, then
\begin{equation*}
\varphi_i(x) = \ell_{g_{ij}(x)} \varphi_j(x)
\end{equation*}
These are the transformation laws for the local decompositions of sections in $\varP$ and $\varE$.

Let $\evF$ be a vector space and $\ell$ a representation (linear action) of $G$. In that case, the associated fiber bundle $\varE$ for the couple $(\evF, \ell)$ is called a vector bundle. The space of smooth sections $\Gamma(\varE)$ is then a $C^\infty(\varM)$-module for the pointwise multiplication: $f(x)s(x)$ for any $f\in C^\infty(\varM)$, $s \in \Gamma(\varE)$ and $x \in \varM$.

Moreover, if $\varE$ and $\varE'$ are vector bundles, then $\varE^\ast$ (dual), $\varE \oplus \varE'$ (Whitney sum), $\varE \otimes \varE'$ (tensor product) and $\exter^\grast \varE$ (exterior product) are defined. They are associated respectively to $(\evF^\ast, \ell^\ast)$, $(\evF \oplus \evF', \ell \oplus \ell')$, $(\evF \otimes \evF', \ell \otimes \ell')$ and $(\exter^\grast \evF, \exter\ell)$. 

Here are now the main examples which will be at the root of the noncommutative geometry introduced in the following, and will permits one to make connections between this noncommutative geometry and the pure geometrical context.

\begin{example}[Tangent and cotangent spaces]
The tangent space $T\varM \rightarrow \varM$, and the cotangent space $T^\ast\varM \rightarrow \varM$ are canonical vector bundles over $\varM$.

The space $\Gamma(T\varM)$ will be denoted by $\Gamma(\varM)$. It is the $C^\infty(\varM)$-module of vector fields on $\varM$. It is also a Lie algebra for the bracket $[X,Y]\cdotaction f = X\cdotaction Y \cdotaction f - Y\cdotaction X \cdotaction f$ for any $f \in C^\infty(\varM)$.

One the other hand, by duality, $\Gamma(T^\ast\varM) = \Omega^1(\varM)$ is the space of $1$-forms on $\varM$. Extending this construction, $\Gamma(\exter^\grast T^\ast\varM) = \Omega^\grast(\varM)$ is the algebra of (de~Rham) differential forms on $\varM$.

If $\varE$ is a vector bundle over $\varM$, then $\Gamma(T^\ast\varM \otimes \varE) = \Omega^\grast(\varM, \varE)$ is the space of differential forms with values in the vector bundle $\varE$, which means that for any $\omega \in \Omega^p(\varM, \varE)$, $X_i \in \Gamma(\varM)$ and $x \in \varM$, $\omega(X_1, \dots, X_p)(x) \in \varE_x$.
\end{example}

\begin{Example}[The endomorphism bundle]
\label{ex-Theendomorphismbundle}
Consider the case where $\evF$ is a finite dimensional vector space. Then $\varE^\ast \otimes \varE$ is associated to $\varP$ for the couple $(\evF^\ast \otimes \evF, \ell^\ast \otimes \ell)$.

One has the identification $\evF^\ast \otimes \evF \simeq \End(\evF)$ by $(\alpha \otimes \varphi)(\varphi') = \alpha(\varphi') \varphi$, where $\End(\evF)$ is the space of endomorphisms of $\evF$.

The vector bundle $\End(\varE) = \varE^\ast \otimes \varE$ is called the endomorphism fiber bundle of $\varE$.

There is a natural pairing $\Gamma(\varE^\ast) \otimes \Gamma(\varE) \rightarrow C^\infty(\varM)$ denoted by $x \mapsto \langle \alpha(x), s(x) \rangle$. One can show that $\Gamma(\varE^\ast \otimes \varE) = \Gamma(\End(\varE))$ is an algebra, which identifies with $\Gamma(\varE^\ast) \otimes_{C^\infty(\varM)} \Gamma(\varE)$ and with the space of $C^\infty(\varM)$-module maps $\Gamma(\varE) \rightarrow \Gamma(\varE)$ by $(\alpha \otimes s)(s')(x) =  \langle \alpha(x), s'(x) \rangle s(x)$.
\end{Example}

\begin{example}[The gauge group and its Lie algebra]
\label{ex-ThegaugegroupanditsLiealgebra}
The group $G$ acts on itself by conjugaison: $\alpha_g(h) = g h g^{-1}$. The associated fiber bundle $\varP \times_\alpha G$ has $G$ as fiber but is not a principal fiber bundle. In particular, this fiber bundle has a global section, defined in any trivialisation by $x \mapsto e$, where $e \in G$ is the unit element. But one knows that the existence of a global section on $\varP$ is equivalent to $\varP$ being trivial.

Denote by $\caG = \Gamma(\varP \times_\alpha G)$ the space of smooth sections. It is a group, called the gauge group of $\varP$: it is the sub-group of vertical automorphisms in $\Aut(\varP)$, the group of all automorphisms of $\varP$. Indeed, any element in $\caG$ is also a $G$-equivariant map $\Phi : \varP \rightarrow G$, which defines the vertical automorphism $p \mapsto p\cdotaction \Phi(p)$. The compatibility condition is ensured by the $G$-equivariance: $p\cdotaction g \mapsto (p\cdotaction g)  \cdotaction \Phi(p\cdotaction g) = (p \cdotaction g) \cdotaction (g^{-1} \Phi(p) g)= (p\cdotaction \Phi(p)) \cdotaction g$.

By construction, one has the short exact sequence of groups:
\begin{equation*}
\xymatrix@1@C=15pt{{\algun} \ar[r] & {\caG} \ar[r] & {\Aut(\varP)} \ar[r] & {\Aut(\varM)} \ar[r] & {\algun}}
\end{equation*}

$G$ acts on the vector space $\kg$ by the adjoint action $\Ad$. Denote by $\Ad \varP = \varP \times_{\Ad} \kg$ the associated vector bundle. The vector space $\Gamma(\Ad \varP)$ is the Lie algebra of the gauge group $\caG$, denoted hereafter by $\lie\caG$.
\end{example}

\subsection{Connections}

Let $\fibre[\pi]{G}{\varP}{\varM}$ be a principal fiber bundle, and let $\varE \rightarrow \varM$ be an associated vector bundle. There is at least three ways to define a connection in this context:
\begin{description}
\item[Geometrical definition:] A connection on $\varP$ is a smooth distribution $H$ in $T\varP$ such that for any $p \in \varP$ and $g \in G$:
\begin{equation*}
T_p\varP = V_p \oplus H_p \quad\text{and}\quad \widetilde{R}_{g \ast} H_p = H_{p\cdotaction g}
\end{equation*}
This defines horizontal vector fields and vertical differential forms (forms which vanish when one of its arguments is horizontal).

One gets the geometrical notion of horizontal lifting of vector fields on $\varM$, which we denote by $\Gamma(\varM) \ni X \mapsto X^h \in \Gamma(\varP)$.

\item[Algebraic definition:] A connection on $\varP$ is a $1$-form on $\varP$ taking values in the Lie algebra $\kg$, $\omega \in \Omega^1(\varP)\otimes \kg$, such that for any $g \in G$ and $X \in \kg$:
\begin{equation*}
\widetilde{R}_{g}^\ast \omega = \Ad_{g^{-1}} \omega \text{ (equivariance)} \quad\text{and}\quad
\omega(X^v) = X \text{ (vertical condition)}
\end{equation*}
The associated horizontal distribution is $H_p = \ker \omega_{|p}$.

\item[Analytic definition:] A connection on $\varE$ is a linear map $\nabla^\varE_X : \Gamma(\varE) \rightarrow \Gamma(\varE)$ defined for any $X \in \Gamma(\varM)$, such that for any $f \in C^\infty(\varM)$, $s \in \Gamma(\varE)$, $X,Y \in \Gamma(\varM)$:
\begin{align*}
\nabla^\varE_X (f s) &= (X\cdotaction f) s + f \nabla^\varE_X s
&
\nabla^\varE_{f X} s &= f \nabla^\varE_X s
&
\nabla^\varE_{X+Y} s &= \nabla^\varE_X s + \nabla^\varE_Y s
\end{align*}
If $s \in \Gamma(\varE)$ corresponds to $\Phi \in \caF_G(\varP,\varF)$, then $\nabla^\varE_X s$ corresponds to $X^h \cdotaction \Phi$.
\end{description}

The equivariance of the connection $1$-form $\omega$ implies the relation
\begin{equation*}
L_{X^v} \omega + [X, \omega] = 0
\end{equation*}
for any $X \in \kg$.

For each of these three definitions, the curvature of a connection can be introduced:
\begin{description}
\item[Geometrical definition:] There exists a geometrical interpretation of the curvature as the obstruction to the closure of horizontal lifts of ``infinitesimal'' closed paths on $\varM$. 

Let $\gamma : [0,1] \mapsto \varM$ be a closed path and let $p \in \varP$. There exists a unique path $\gamma^h : [0,1] \mapsto \varP$ such that $\gamma^h(0)=p$ and $\dot{\gamma}^h(t) \in H_{\gamma^h(t)}$ for any $t  \in [0,1]$. $\gamma^h$ is a horizontal lifting of $\gamma$. One has $\gamma^h(1) \neq \gamma^h(0)=p$ \textsl{a priori}, but they are in the same fiber, so that the deficiency is in $G$. 

When the path $\gamma$ is shrunk to an infinitesimal path, the deficiency is an element in $\kg$ which depends only on $\dot{\gamma}(0)$ and $\dot{\gamma}(1)$. This is the curvature.

\item[Algebraic definition:] The curvature is the equivariant horizontal $2$-form $\Omega \in \Omega^2(\varP) \otimes \kg$ defined for any $\caX, \caY \in \Gamma(\varP)$ by
\begin{equation*}
\Omega(\caX, \caY) = \dd\omega(\caX, \caY) + [\omega(\caX), \omega(\caY)]
\end{equation*}
It satisfies the Bianchi identity 
\begin{equation*}
\dd \Omega + [\omega, \Omega] = 0
\end{equation*}

\item[Analytic definition:] Given $\nabla^\varE_X : \Gamma(\varE) \rightarrow \Gamma(\varE)$, the curvature $R^\varE(X,Y)$ is the map defined for any $X,Y \in \Gamma(\varM)$ by
\begin{equation*}
R^\varE(X,Y) = \nabla^\varE_X \nabla^\varE_Y - \nabla^\varE_Y \nabla^\varE_X - \nabla^\varE_{[X,Y]} : \Gamma(\varE) \rightarrow \Gamma(\varE)
\end{equation*}
The remarkable fact is that this particular combinaison is a $C^\infty(\varM)$-module map.
\end{description}

One can connect these definitions by the following relations. Let $\eta$ be the representation of $\kg$ on $\evF$ induced by the representation $\ell$ of $G$. If $s \in \Gamma(\varE)$ corresponds to $\Phi \in \caF_G(\varP,\varF)$, then $R^\varE(X,Y) s$ corresponds to $\eta(\Omega(\caX, \caY)) \cdotaction \Phi$ for any $\caX, \caY$ such that $\pi_\ast \caX = X$ and $\pi_\ast \caY = Y$.

Let $\omega \in \Omega^1(\varP)\otimes \kg$ be a connection $1$-form on $\varP$, and $\Omega$ its curvature. Let $(U, \phi)$ be local trivialisation of $\varP$, and $s$ its associated local section.

One can define the local expression of the connection and the curvature in this trivialisation as the pull-back of $\omega$ and $\Omega$ by $s : U \rightarrow \varP$:
\begin{align*}
A &= s^\ast \omega \in \Omega^1(U)\otimes \kg
&
F &= s^\ast \Omega \in \Omega^2(U)\otimes \kg
\end{align*}
If $(U_i, \phi_i)$ and $(U_j, \phi_j)$ are two local trivialisations, on $U_i \cap U_j \neq \ensvide$ one has the well-known relations
\begin{align}
\label{eq-gluingrelationconnectioncurvature}
A_j &=  g^{-1}_{ij} A_i g_{ij} + g^{-1}_{ij} \dd g_{ij}
&
F_j &=  g^{-1}_{ij} F_i g_{ij}
\end{align}
with obvious notations. A family of $1$-forms $\{ A_i\}_i$ satisfying these gluing relations defines a connection $1$-form on $\varP$. This is (too) often used in the physical literature as a possible definition of a connection and its curvature.

\begin{remark}[Intermediate construction]
\label{rmk-intermediateconstruction}
\begin{table}
\centering
\begin{tabular}{p{0.35\textwidth}|p{0.55\textwidth}}
\textbf{\mathversion{bold}Globally on $\varP$}
&
\textbf{\mathversion{bold}Locally on $\varM$}\\
\hline
$\omega \in \Omega^1(\varP)\otimes \kg$, equivariant, vertical condition.
&
Family of local $1$-forms $\{ A_i\}_i$, $A_i \in \Omega^1(U_i)\otimes \kg$, satisfying gluing non homogeneous relations.\\
$\Omega \in \Omega^2(\varP)\otimes \kg$, equivariant and horizontal.
&
Family of local $2$-forms $\{ F_i\}_i$, $F_i \in \Omega^2(U_i)\otimes \kg$, satisfying gluing homogeneous relations.
\end{tabular}
\caption{The two ordinary constructions of the connections and curvature, the global one on $\varP$ and the local one on $\varM$.}
\label{tab-topanddownconstruction}
\end{table}

We summarize in Table~\ref{tab-topanddownconstruction} the two common ways to introduce a connection as differential objects, either as a global $1$-form on $\varP$ or as a family of local $1$-forms on $\varM$.

It is well known that, using the homogeneous gluing relations for the $F_i$'s, or using the equivariant and horizontal property of $\Omega$, one can show that the curvature is also a section of the associated vector bundle $\exter^2 T^\ast \varM \otimes \Ad \varP$, \textsl{i.e.} a global $2$-forms on $\varM$ with values in the vector bundle $\Ad \varP = \varP \times_{\Ad} \kg$. We denote by $\gF \in \Omega^2(\varM, \Ad \varP)$ this $2$-form.

Because of the inhomogeneous gluing relations for the $A_i$'s, the connection cannot be the section of such an ``intermediate'' construction between forms on $\varP$ and local forms on the $U_i$'s.

Let us mention here that in the noncommutative geometry introduced in the following, this intermediate construction is possible also for the connection $1$-form. See Remark~\ref{rmk-Theintermediateconstructioninordinarygeometry}.
\end{remark}

\subsection{Gauge transformations}

We saw that the gauge group $\caG = \Gamma(\varP \times_\alpha G)$ acts on $\varP$. To any $a \in \caG$ one can associate a $G$-equivariant map $\Phi  : \varP \rightarrow G$. The corresponding vertical diffeomorphism $\varP \rightarrow \varP$ defined by $a$ is also denoted by $a$

Let $\omega \in \Omega^1(\varP)\otimes \kg$ be a connection on $\varP$. Then one can show that the pull-back $a^\ast \omega$ is also a connection and $a^\ast \Omega$ is its curvature. Explicitly, one can establish the formulae $a^\ast \omega = \Phi^{-1} \omega \Phi + \Phi^{-1} \dd \Phi$ and $a^\ast \Omega = \Phi^{-1} \Omega \Phi$, which look very similar to \eqref{eq-gluingrelationconnectioncurvature}, but are not the same: here we perform some active transformation on the space of connections while in \eqref{eq-gluingrelationconnectioncurvature} we look at the same connection in different trivialisations. This is the difference between active and passive transformation laws.

In order to get the action of the Lie algebra of the gauge group, consider $\Phi = \exp(\xi)$ with $\xi : \varP \rightarrow \kg$, $G$-equivariant, so that $\xi$ defines an element in $\lie\caG=\Gamma(\Ad \varP)$. Then the infinitesimal action on connections and curvatures take the form:
\begin{align*}
\omega &\mapsto \dd \xi + [\omega, \xi]
&
\Omega &\mapsto [\Omega, \xi]
\end{align*}

\section{Derivation-based noncommutative geometry}

In this section, we introduce the algebraic context in which the noncommutative geometry we are interested in is constructed. The differential calculus we consider here has been introduced in \cite{DuVi:88} and has been exposed and studied for various algebras, for instance in \cite{DuViKernMado:90a}, \cite{DuViKernMado:90b}, \cite{Mass:07}, \cite{Mass:11}, \cite{Mass:14},\cite{Mass:15}, \cite{DuViMich:94}, \cite{DuViMich:96}, \cite{DuViMich:97}.

\subsection{Derivation-based differential calculus}

Let $\algA$ be an associative algebra with unit $\bbbone$. Denote by $\caZ(\algA)$ the center of $\algA$.

\begin{definition}[Vector space of derivations of $\algA$]
The vector space of derivations of $\algA$ is the space
$\der(\algA) = \{ \kX : \algA \rightarrow \algA \ / \ \kX \text{ linear}, \kX(ab) = \kX(a) b + a \kX(b), \forall a,b\in \algA\}$
\end{definition} 

The essential properties of this space are contained in the following:
\begin{proposition}[Structure of $\der(\algA)$]
$\der(\algA)$ is a Lie algebra for the bracket $[\kX, \kY ]a = \kX  \kY a - \kY \kX a$ ($\forall \kX,\kY \in \der(\algA)$) and a $\caZ(\algA)$-module for the product $(f\kX )a = f(\kX a)$ ($\forall f \in \caZ(\algA)$, $\forall \kX \in \der(\algA)$).

The subspace $\Int(\algA) = \{ \adrep_a : b \mapsto [a,b]\ / \ a \in \algA\} \subset \der(\algA)$, called the vector space of inner derivations, is a Lie ideal and a $\caZ(\algA)$-submodule.

With $\Out(\algA)=\der(\algA)/\Int(\algA)$, there is a short exact sequence of Lie algebras and $\caZ(\algA)$-modules
\begin{equation}
\label{eq-secderivations}
\xymatrix@1@C=15pt{{\algzero} \ar[r] & {\Int(\algA)} \ar[r] & {\der(\algA)} \ar[r] & {\Out(\algA)} \ar[r] & {\algzero}}
\end{equation}
\end{proposition} 

In case $\algA$ has an involution $a \mapsto a^\ast$, one can define real derivations:
\begin{definition}[Real derivations for involutive algebras]
If $\algA$ is an involutive algebra, the derivation $\kX \in \der(\algA)$ is real if $(\kX a)^\ast = \kX a^\ast$ for any $a\in \algA$. We denote by $\der_\gR(\algA)$ the space of real derivations.
\end{definition}

\begin{definition}[The graded differential algebra $\underline{\Omega}^\grast_\der(\algA)$]
Let $\underline{\Omega}^n_\der(\algA)$ be the set of $\caZ(\algA)$-multilinear antisymmetric maps from $\der(\algA)^n$ to $\algA$, with $\underline{\Omega}^0_\der(\algA) = \algA$, and let
\begin{equation*}
\underline{\Omega}^\grast_\der(\algA) =\textstyle \bigoplus_{n \geq 0} \underline{\Omega}^n_\der(\algA)
\end{equation*}
We introduce on $\underline{\Omega}^\grast_\der(\algA)$ a structure of $\gN$-graded differential algebra using the product
\begin{equation*}
(\omega\eta)(\kX_1, \dots, \kX_{p+q}) =
 \frac{1}{p!q!} \sum_{\sigma\in \kS_{p+q}} (-1)^{\sign(\sigma)} \omega(\kX_{\sigma(1)}, \dots, \kX_{\sigma(p)}) \eta(\kX_{\sigma(p+1)}, \dots, \kX_{\sigma(p+q)})
\end{equation*}
and using the differential $\dd$ (of degree $1$) defined by the Koszul formula
\begin{multline*}
\dd\omega(\kX_1, \dots , \kX_{n+1}) = \sum_{i=1}^{n+1} (-1)^{i+1} \kX_i \omega( \kX_1, \dots \omi{i} \dots, \kX_{n+1}) \\[-5pt]
 + \sum_{1\leq i < j \leq n+1} (-1)^{i+j} \omega( [\kX_i, \kX_j], \dots \omi{i} \dots \omi{j} \dots , \kX_{n+1}) 
\end{multline*}
\end{definition}

\begin{definition}[The graded differential algebra $\Omega^\grast_\der(\algA)$]
Denote by $\Omega^\grast_\der(\algA) \subset \underline{\Omega}^\grast_\der(\algA)$ the sub differential graded algebra generated in degree $0$ by $\algA$.
\end{definition} 

Notice that by definition, every element in $\Omega^n_\der(\algA)$ is a sum of terms of the form $a_0 \dd a_1 \cdots \dd a_n$ for $a_0, \dots, a_n \in \algA$.

\bigskip
The previous definitions are motivated by the following important example which shows that these definitions are correct generalisations of the space of ordinary differential forms on a manifold:
\begin{example}[The algebra $\algA = C^\infty(\varM)$]
Let $\varM$ be a smooth manifold and let $\algA = C^\infty(\varM)$. The center of this algebra is $\algA$ itself: $\caZ(\algA) = C^\infty(\varM)$. The Lie algebra of derivations is exactly the Lie algebra of smooth vector fields on $\varM$: $\der(\algA) = \Gamma(\varM)$. In that case, there is no inner derivations, $\Int(\algA) = \algzero$, so that $\Out(\algA) = \Gamma(\varM)$.

The two graded differential algebras coincide with the graded differential algebra of de~Rham forms on $\varM$: $\Omega^\grast_\der(\algA) = \underline{\Omega}^\grast_\der(\algA) = \Omega^\grast(\varM)$
\end{example}

In the previous definitions of the graded differential calculi, one is not bounded to consider the entire Lie algebra of derivations:
\begin{definition}[Restricted derivation-based differential calculus]
Let $\kg \subset \der(\algA)$ be a sub Lie algebra and a sub $\caZ(\algA)$-module. The restricted derivation-based differential calculus $\underline{\Omega}^\grast_\kg(\algA)$ associated to $\kg$ is defined as the set of $\caZ(\algA)$-multilinear antisymmetric maps from $\kg^n$ to $\algA$ for $n \geq 0$, using the previous formulae for the product and the differential.
\end{definition} 

\bigskip
Now, let $\kg$ be any Lie subalgebra of $\der(\algA)$. Then $\kg$ defines a natural operation in the sense of H.~Cartan on the graded differential algebra $(\underline{\Omega}^\grast_\der(\algA), \dd)$. The interior product is the graded derivation of degree $-1$ on $\underline{\Omega}^\grast_\der(\algA)$ defined by
\begin{align*}
i_\kX : \underline{\Omega}^n_\der(\algA)  &\rightarrow \underline{\Omega}^{n-1}_\der(\algA)
&
(i_\kX \omega)( \kX_1, \dots , \kX_{n-1})  = \omega (\kX, \kX_1, \dots , \kX_{n-1})
\end{align*}
$\forall \kX\in \kg$, $\forall \omega \in \underline{\Omega}^n_\der(\algA)$ and $ \forall \kX_i \in \der(\algA)$. By definition, $i_\kX$ is $0$ on $\underline{\Omega}^0_\der(\algA)=\algA$.

The associated Lie derivative is the graded derivation of degree $0$ on $\underline{\Omega}^\grast_\der(\algA)$ given by
\begin{equation*}
L_\kX = i_\kX \dd + \dd i_\kX  : \underline{\Omega}^n_\der(\algA) \rightarrow \underline{\Omega}^{n}_\der(\algA)
\end{equation*}

One can easily verify the relations defining a Cartan operation:
\begin{align*}
i_\kX i_\kY + i_\kY i_\kX &= 0 & L_\kX i_\kY - i_\kY L_\kX &= i_{[\kX,\kY]} \\
L_\kX L_\kY - L_\kY L_\kX &= L_{[\kX,\kY]} & L_\kX \dd - \dd L_\kX &= 0
\end{align*}

One can then associate to this operation the following subspaces of $\underline{\Omega}^\grast_\der(\algA)$:
\begin{itemize}
\item The horizontal subspace is the kernel of all the $i_\kX$ for $\kX\in \kg$. This is a graded algebra.

\item The invariant subspace is the kernel of all the $L_\kX$ for $\kX\in \kg$. This is a graded differential algebra.

\item The basic subspace is the kernel of all the $i_\kX$ and $L_\kX$ for $\kX\in \kg$. This is a graded differential algebra.
\end{itemize}

For instance, $\kg = \Int(\algA)$ defines such an operation.

\subsection{Noncommutative connections and their properties}

Noncommutative connections play a central role in noncommutative differential geometry. They are all based on some generalisation of what we called the analytic definition of ordinary connections, where one replaces the $C^\infty$-module of sections of a vector bundle by a more general (finitely projective) module over the algebra. Various definitions has been proposed, for instance to take into account some bimodule structures. Here we only consider right $\algA$-modules.

\subsubsection{Definitions and general properties}

Let $\modM$ be a right $\algA$-module.

\begin{definition}[Noncommutative connections, curvature]
A noncommutative connection on $\modM$ for the differential calculus based on derivations is a linear map $\widehat{\nabla}_\kX : \modM \rightarrow \modM$, defined for any $\kX \in \der(\algA)$, such that $\forall \kX,\kY \in \der(\algA)$, $\forall a \in \algA$, $\forall m \in \modM$, $\forall f \in \caZ(\algA)$:
\begin{align*}
\widehat{\nabla}_\kX (m a) &= m(\kX a) + (\widehat{\nabla}_\kX m) a,
&
\widehat{\nabla}_{f\kX} m &= f \widehat{\nabla}_\kX m,
&
\widehat{\nabla}_{\kX + \kY} m &= \widehat{\nabla}_\kX m + \widehat{\nabla}_\kY m
\end{align*}
The curvature of $\widehat{\nabla}$ is the linear map $\hR(\kX, \kY) : \modM \rightarrow \modM$ defined for any $\kX, \kY \in \der(\algA)$ by
\begin{equation*}
\hR(\kX, \kY) m = [ \widehat{\nabla}_\kX, \widehat{\nabla}_\kY ] m - \widehat{\nabla}_{[\kX, \kY]}m
\end{equation*}
\end{definition} 

This definition is an adaptation to the derivation-based noncommutative calculus of the ordinary (analytic) definition of connections. Notice that we have to make use of the center $\caZ(\algA)$ of the algebra $\algA$ for one of the above relations, which means that we have to differentiate the respective roles of the algebra and of its center. 

\begin{proposition}[General properties]
The space of connections is an affine space modeled over the vector space $\Hom^{\algA}(\underline{\Omega}^1_\der(\algA), \modM)$ (right $\algA$-module morphisms from $\underline{\Omega}^1_\der(\algA)$ to $\modM$).

$\hR(\kX, \kY) : \modM \rightarrow \modM$ is a right $\algA$-module morphism.
\end{proposition} 

\begin{definition}[The gauge group]
The gauge group of $\modM$ is the group of automorphisms of $\modM$ as a right $\algA$-module.
\end{definition} 

\begin{proposition}[Gauge transformations]
For any $\Phi$ in the gauge group of $\modM$ and any noncommutative connection $\widehat{\nabla}$, the map $\widehat{\nabla}^\Phi_\kX = \Phi^{-1}\circ \widehat{\nabla}_\kX \circ \Phi : \modM \rightarrow \modM$ is a noncommutative connection.

This defines the action of the gauge group on the space of noncommutative connections.
\end{proposition} 


Suppose now that $\algA$ is an involutive algebra and let as before $\modM$ be a right $\algA$-module.

\begin{definition}[Hermitean structure, compatible noncommutative connections]
A Hermitean structure on $\modM$ is a sesquilinear form $\langle-,-\rangle:\modM \times \modM \rightarrow \algA$ such that, $\forall m_1, m_2 \in \modM$, $\forall a_1, a_2 \in \algA$,
\begin{align*}
\langle m_1, m_2\rangle^\ast &= \langle m_2, m_1\rangle
&
\langle m_1 a_1, m_2 a_2\rangle &= a_1^\ast \langle m_1, m_2\rangle a_2
\end{align*}
A noncommutative connection $\widehat{\nabla}$ is compatible with $\langle-,-\rangle$ if, $\forall m_1, m_2 \in \modM$, $\forall \kX \in \der_\gR(\algA)$,
\begin{equation*}
\kX \langle m_1, m_2\rangle = \langle \widehat{\nabla}_\kX m_1, m_2\rangle + \langle m_1, \widehat{\nabla}_\kX m_2\rangle
\end{equation*}
\end{definition} 

\begin{definition}[``Unitary'' gauge transformations]
An element $\Phi$ in the gauge group is compatible with the Hermitean structure if, for any $m_1, m_2 \in \modM$, one has $\langle \Phi(m_1), \Phi(m_2)\rangle = \langle m_1, m_2\rangle$. In that case, we refer to such a gauge transformation as a ``unitary'' gauge transformation.
\end{definition} 

\begin{lemma}
The space of compatible noncommutative connections with $\langle-,-\rangle$ is stable under ``unitary'' gauge transformations.
\end{lemma}

\subsubsection{The right $\algA$-module $\modM=\algA$}

As a special case of the previous general situation, we consider the right $\algA$-module $\modM=\algA$. Let $\widehat{\nabla}_\kX : \algA \rightarrow \algA$ be a noncommutative connection.

\begin{proposition}[Noncommutative connections on $\modM=\algA$]
\label{prop-NoncommutativeconnectionsonmodM=algA}
The noncommutative connection $\widehat{\nabla}$ is completely determined by $\widehat{\nabla}_\kX \bbbone = \omega(\kX)$, with $\omega \in \underline{\Omega}^1_\der(\algA)$, by the relation 
\begin{equation*}
\widehat{\nabla}_\kX a = \kX a + \omega(\kX) a
\end{equation*}
The curvature of $\widehat{\nabla}$ is the multiplication on the left on $\algA$ by the noncommutative $2$-form 
\begin{equation*}
\Omega(\kX, \kY) = \dd\omega (\kX, \kY) + [ \omega(\kX), \omega(\kY) ]
\end{equation*}
The gauge group is identified with the invertible elements $g \in \algA$ by $\Phi_g(a) = ga$ and the gauge transformations on $\widehat{\nabla}$ take the following form on $\omega$ and $\Omega$:
\begin{align*}
\omega &\mapsto \omega^g = g^{-1} \omega g + g^{-1} \dd g
&
\Omega &\mapsto \Omega^g = g^{-1} \Omega g
\end{align*}
$\widehat{\nabla}^0_\kX$, defined by $a \mapsto \kX a$, is a noncommutative connection on $\algA$.
\end{proposition} 

The gauge transformations on the noncommutative forms $\omega$ and $\Omega$ are clearly of the same nature as the one encountered in ordinary differential geometry. Nevertheless, the relations are different: the differential operator is the noncommutative differential here.

In the particular case when $\algA$ is involutive, one can define a canonical Hermitean structure on $\modM$ by $\langle a, b\rangle = a^\ast b$. Then, $U(\algA) = \{ u \in \algA\ / \ u^\ast u = u u^\ast = \bbbone\}$, the group of unitary elements of $\algA$, identifies with the unitary gauge group.

\bigskip
Let us stress the following important point.
\begin{remark}[Vector space \textit{versus} gauge transformations]
We saw that the space of noncommutative connections is an affine space, but here it looks like the vector space $\underline{\Omega}^1_\der(\algA)$. In fact, one can show that gauge transformations are not compatible with this linear structure:
\begin{align*}
(\lambda_1 \omega_1 + \lambda_2 \omega_2)^u &= u^{-1} (\lambda_1 \omega_1 + \lambda_2 \omega_2) u + u^{-1} \dd u \\
\lambda_1 \omega_1^u + \lambda_2 \omega_2^u & = \lambda_1 (u^{-1} \omega_1 u + u^{-1} \dd u) + \lambda_2 (u^{-1} \omega_2 u + u^{-1} \dd u)
\end{align*}
are not equal except for $\lambda_1 + \lambda_2 = 1$.
\end{remark} 

The following proposition applies in some important examples:
\begin{proposition}[Canonical gauge invariant noncommutative connection]
\label{Canonicalgaugeinvariantnoncommutativeconnection}
If there exists a noncommutative $1$-form $\xi \in \underline{\Omega}^1_\der(\algA)$ such that $\dd a = [\xi, a]$ for any $a \in \algA$, then the canonical noncommutative connection defined by $\widehat{\nabla}^{-\xi}_\kX a = \kX a - \xi(\kX) a$ can be written as $\widehat{\nabla}^{-\xi}_\kX a = - a \xi(\kX)$. 

Moreover, this canonical noncommutative connection is gauge invariant.
\end{proposition} 

\begin{proof}
One has $\kX a = [\xi(\kX), a]$, so that $\widehat{\nabla}^{-\xi}_\kX a = [\xi(\kX), a] - \xi(\kX) a = - a \xi(\kX)$. 

Let $u \in U(\algA)$ be a unitary gauge transformation. Its action on the noncommutative $1$-form $-\xi$ is $(-\xi)^u = - u^{-1} \xi u + u^{-1} \dd u = u^{-1}(-\xi u + [\xi, u]) = u^{-1}(- u \xi) =  -\xi$, which shows that this noncommutative connection is indeed gauge invariant.
\end{proof}

As can be immediately seen, this situation can't occur in the commutative case (ordinary differential geometry) because for any $1$-form $\xi$, one has $[\xi, a]=0$. Below, we will encounter a situation where such a noncommutative $1$-form exists, in the context of the algebra $M_n(\gC)$ of complex matrices. An other important example where such an invariant noncommutative connection makes its appearance is the Moyal algebra. These two examples share in common that they only have inner derivations. They are highly noncommutative situation in this respect, even if the Moyal algebra can be considered as a deformation of some commutative algebra of ordinary smooth functions.

\subsubsection{The right $\algA$-module $\modM=\algA^N$}

As an other special case of right $\algA$-modules, we consider now the case where the right $\algA$-module is $\modM=\algA^N$. Denote by $e_i = (0, \dots, \bbbone, \dots, 0)$, for $i=1,\dots, N$, a canonical basis of the right module $\algA^N$. We look at $m= e_i a^i \in \modM$ as a column vector for the $a^i$'s, so that we use some matrix product notations. We also use the notation $\kX m = e_i (\kX a^i)$ for any derivation $\kX$ of $\algA$.

Let $\widehat{\nabla}_\kX : \algA^N \rightarrow \algA^N$ be a noncommutative connection. 

\begin{proposition}[Noncommutative connections on $\modM=\algA^N$]
\label{NoncommutativeconnectionsonmodMalgAN}
$\widehat{\nabla}$ is completely determined by $N^2$ noncommutative $1$-forms $\omega_{i}^j \in \underline{\Omega}^1_\der(\algA)$ defined by $\widehat{\nabla}_\kX e_i = e_j \omega_{i}^j(\kX)$, through the relation $\widehat{\nabla}_\kX m = \kX m + \omega(\kX) m$, with $\omega = (\omega_{i}^j) \in M_N(\underline{\Omega}^1_\der(\algA))$. 

The curvature of $\widehat{\nabla}$ is the multiplication on the left on $\algA^N$ by the matrix of noncommutative $2$-forms $\Omega = \dd\omega + [ \omega, \omega ] \in M_N(\underline{\Omega}^2_\der(\algA))$.

The gauge group of $\algA^N$ is $GL_N(\algA)$ (invertibles in $M_N(\algA)$), which acts by left (matrix) multiplication. The gauge transformations take the forms $\omega^g = g^{-1} \omega g + g^{-1} \dd g$ and $\Omega^g = g^{-1} \Omega g$ in matrix notations.

$\widehat{\nabla}^0_\kX$, defined by $m \mapsto \kX m$, is a noncommutative connection on $\algA^N$.
\end{proposition} 

In the particular case when $\algA$ is involutive, the natural Hermitean structure on $\modM$ is defined by $\langle (a^i), (b^j)\rangle = \sum_{i=1}^N (a^i)^\ast b^i$. Then, $U_N(\algA) = \{ u \in M_N(\algA)\ / \ u^\ast u = u u^\ast = \bbbone_N\}$, the group of unitary elements of $M_N(\algA)$, is the unitary gauge group.

\subsubsection{The projective finitely generated right $\algA$-modules}

From the Serre-Swan theorem, one knows that any vector bundle on a smooth manifold $\varM$ is characterised by its space of smooth sections as a projective finitely generated right module (p.f.g.m.) over $C^\infty(\varM)$. The natural generalisation of vector bundles in noncommutative geometry is then taken to be the projective finitely generated right $\algA$-modules.

Let $\modM$ be such a projective finitely generated right $\algA$-modules. $\modM$ is a direct summand in $\algA^N$, so that there exists a projection $p \in M_N(\algA)$ such that $\modM = p \algA^N$.

\begin{proposition}[Noncommutative connections on p.f.g.m.]
If $\widehat{\nabla}$ is a noncommutative connection on the right $\algA$-module $\algA^N$, then $m \mapsto p \widehat{\nabla}_\kX m$ defines a noncommutative connection on $\modM$, where $m \in \modM \subset \algA^N$.

The curvature of the noncommutative connection obtained this way from the canonical noncommutative connection $\widehat{\nabla}^0_\kX$ of Proposition~\ref{NoncommutativeconnectionsonmodMalgAN}, is the multiplication on the left on $\modM \subset \algA^N$ by the matrix of noncommutative $2$-forms $p \dd p \dd p$.
\end{proposition} 

\begin{example}[The algebra $\algA = C^\infty(\varM)$]
We saw that the noncommutative derivation-based differential calculus is the ordinary de~Rham calculus. Using the equivalence given in the Serre-Swan theorem, the definitions of (ordinary) connections and of noncommutative connections coincide.
\end{example}

\subsection{Two important examples}

\subsubsection{The algebra $\algA = M_n(\gC) = M_n$}

Let us consider the case $\algA = M_n(\gC) = M_n$, the finite dimensional algebra of $n\times n$ complex matrices. This is an involutive algebra for the adjointness of matrices.

First, we summarize the general properties of its derivation-based differential calculus, which is described in \cite{DuVi:88}, \cite{DuViKernMado:90a} and \cite{Mass:11}.
\begin{proposition}[General properties of the differential calculus]
One has the following results:
\begin{itemize}
\item $\caZ(M_n) = \gC$.

\item $\der(M_n) = \Int(M_n) \simeq \ksl_n =\ksl(n,\gC)$ (traceless matrices). The explicit isomorphism associates to any $\gamma \in \ksl_n(\gC)$ the derivation $\adrep_\gamma : a \mapsto [\gamma, a]$. 

$\der_\gR(M_n) = \ksu(n)$ and $\Out(M_n) = \algzero$.

\item $\underline{\Omega}^\grast_\der(M_n) = \Omega^\grast_\der(M_n) \simeq M_n \otimes \exter^\grast \ksl_n^\ast$, with the differential $\dd'$ coming from the differential of the differential complex of the Lie algebra $\ksl_n$ represented on $M_n$ by the adjoint representation (commutator).

\item There exits a canonical noncommutative $1$-form $i\theta \in \Omega^1_\der(M_n)$ such that  for any $\gamma \in M_n(\gC)$
\begin{equation*}
i\theta(\adrep_{\gamma}) = \textstyle\gamma - \frac{1}{n} \tr (\gamma)\bbbone
\end{equation*}
This noncommutative $1$-form $i\theta$ makes the explicit isomorphism $\Int(M_n(\gC)) \xrightarrow{\simeq} \ksl_n$.

\item $i\theta$ satisfies the relation $\dd' (i\theta) - (i\theta)^2 = 0$. This makes $i\theta$ look very much like the Maurer-Cartan form in the geometry of Lie groups (here $SL_n(\gC)$).

\item For any $a \in M_n$, one has $\dd' a = [i\theta, a] \in \Omega^1_\der(M_n)$. This relation is no more true in higher degrees.
\end{itemize}

\end{proposition} 

Let us now introduce a particular basis of this algebra, which permits one to perform explicit computations. Denote by $\{E_k\}_{k=1, \dots, n^2-1}$ a basis for $\ksl_n$ of hermitean matrices.  Then, it defines a basis for the Lie algebra $\der(M_n) \simeq \ksl_n$ through the $n^2-1$ derivations $\partial_k = ad_{i E_k}$, which are real derivations. Adjoining the unit $\bbbone$ to the $E_k$'s, one gets a basis for $M_n$.

Let us define the $\theta^\ell$'s in $\ksl_n^\ast$ by duality: $\theta^\ell(\partial_k) = \delta^\ell_k$. Then $\{\theta^\ell\}$ is a basis of $1$-forms in $\exter^\grast \ksl_n^\ast$. By definition, they anticommute: $\theta^\ell \theta^k = - \theta^k \theta^\ell$ in this exterior algebra.

Define the structure constants by $[E_k, E_\ell] = C^m_{k \ell} E_m$. Then one can show that the differential $\dd'$ takes the explicit form:
\begin{align*}
\dd' \bbbone &= 0 & \dd' E_k &= -C^m_{k\ell} E_m \theta^\ell & \dd' \theta^k &= -\frac{1}{2} C^k_{\ell m} \theta^\ell \theta^m
\end{align*}

The noncommutative $1$-form $i\theta$ can be written as $i\theta = i E_k \theta^k \in M_n \otimes \exter^1 \ksl_n^\ast$. It is obviously independent of the chosen basis.

\begin{proposition}[The cohomology of the differential calculus]
The cohomology of the differential algebra $(\Omega^\grast_\der(M_n), \dd')$ is
\begin{equation*}
H^\grast( \Omega^\grast_\der(M_n), d') = \caI(\exter^\grast\ksl_n^\ast)
\end{equation*}
the algebra of invariant elements for the natural Lie derivative. 

Recall that the algebra $\caI(\exter^\grast\ksl_n^\ast)$ is the graded commutative algebra generated by elements $c^n_{2 r - 1}$ in degree $2r-1$ for $r \in \{ 2,3, \dots, n\}$.
\end{proposition}

Let us introduce the symmetric matrix $g_{k \ell} = \frac{1}{n} \tr(E_k E_\ell)$. Then the $g_{k \ell}$'s define a natural metric (scalar product) on $\der(M_n)$ with the relation $g(\partial_k, \partial_\ell) = g_{k \ell}$.

Now, one can show that every differential form of maximal degree $\omega \in \Omega^{n^2-1}_\der(M_n)$ can be written uniquely in the form
\begin{equation*}
\omega = a \sqrt{|g|} \theta^1 \cdots \theta^{n^2-1}
\end{equation*}
where $a \in M_n$ and where $|g|$ is the determinant of the matrix $(g_{k \ell})$. 

\begin{definition}[Noncommutative integration]
One defines a noncommutative integration 
\begin{equation*}
\int_{\text{n.c.}} : \Omega^\grast_\der(M_n) \rightarrow \gC
\end{equation*}
by $\int_{\text{n.c.}} \omega = \frac{1}{n} \tr(a)$ if $\omega \in \Omega^{n^2-1}_\der(M_n)$ written as above, and $0$ otherwise.

This integration satisfies the closure relation
\begin{equation*}
\int_{\text{n.c.}} \dd' \omega = 0
\end{equation*}
\end{definition} 

Let us now consider the right $\algA$-module $\modM = \algA$. 

The noncommutative $1$-form $-i\theta$ defines a canonical noncommutative connection by the relation $\widehat{\nabla}^{-i\theta}_\kX a = \kX a - i\theta(\kX) a$ for any $a \in \algA$. 

\begin{proposition}[Properties of $\widehat{\nabla}^{-i\theta}$]
For any $a \in M_n$ and $\kX = \adrep_\gamma \in \der(M_n)$ (with $\tr \gamma = 0$), one has 
\begin{equation*}
\widehat{\nabla}^{-i\theta}_\kX a = - a i\theta(\kX) = - a \gamma
\end{equation*}
$\widehat{\nabla}^{-i\theta}$ is gauge invariant.

The curvature of the noncommutative connection $\widehat{\nabla}^{-i\theta}$ is zero.
\end{proposition} 

\begin{proof}
This is a consequence of the existence of the canonical gauge invariant noncommutative connection implied by the relation $\dd' a = [i\theta, a]$ (Proposition~\ref{Canonicalgaugeinvariantnoncommutativeconnection}).

The curvature is the noncommutative $2$-form $\Omega(\kX, \kY) = \dd'i\theta (\kX, \kY) + [ i\theta(\kX), i\theta(\kY) ] = \dd'i\theta (\kX, \kY) + (i\theta)^2(\kX, \kY) = 0$.
\end{proof}

Let us now consider the right $\algA$-module $\modM = M_{r,n}$, the vector space of $r\times n$ complex matrices with the obvious right module structure and the Hermitean structure $\langle m_1, m_2 \rangle = m_1^\ast m_2 \in M_n$.

\begin{proposition}[$\widehat{\nabla}^{-i\theta}$, flat noncommutative connections]
The noncommutative connection $\widehat{\nabla}^{-i\theta}_\kX m = - m i\theta(\kX)$ is well defined, it is compatible with the Hermitean structure and its curvature is zero.

Any noncommutative connection can be written $\widehat{\nabla}_\kX a = \widehat{\nabla}^{-i\theta}_\kX a + A(\kX) a$ for $A = A_k \theta^k$ with $A_k \in M_r$. The curvature of $\widehat{\nabla}$ is the multiplication on the left by the $M_r$-valued noncommutative $2$-form
\begin{equation*}
F = \frac{1}{2}( [A_k, A_\ell] - C^m_{k \ell} A_m) \theta^k \theta^\ell
\end{equation*}
This curvature vanishes if and only if $A : \ksl_n \rightarrow M_r$ is a representation of the Lie algebra $\ksl_n$.

Two flat connections are in the same gauge orbit if and only if the corresponding Lie algebra representations are equivalent.
\end{proposition} 

For the proof, we refer to \cite{DuViKernMado:90a}.

\subsubsection{The algebra $\algA = C^\infty(\varM) \otimes M_n$}
\label{subsubsection-ThealgebraalgA=Cinfty(varM)otimesMn}

As a second important example, we consider now the mixed of the two algebras $C^\infty(\varM)$ and $M_n(\gC)$ studied before, in the form of matrix valued functions on a smooth manifold $\varM$ ($\dim \varM = m$).

The derivation-based differential calculus for this algebra was first considered in \cite{DuViKernMado:90b}:
\begin{proposition}[General properties of the differential calculus]
\label{prop-Generalpropertiesofthedifferentialcalculustrivialcase}
One has the following results:
\begin{itemize}
\item $\caZ(\algA) = C^\infty(\varM)$.
\item $\der(\algA) = [\der(C^\infty(\varM))\otimes \bbbone ] \oplus [ C^\infty(\varM) \otimes \der(M_n) ] = \Gamma(\varM) \oplus [C^\infty(\varM) \otimes \ksl_n]$ as Lie algebras and $C^\infty(\varM)$-modules. In the following, we will use the notations: $\kX = X + \adrep_\gamma$ with $X \in \Gamma(\varM)$ and $\gamma \in C^\infty(\varM) \otimes \ksl_n = \algA_0$ (traceless elements in $\algA$).

One can identify $\Int(\algA) = \algA_0$ and $\Out(\algA) = \Gamma(\varM)$.

\item $\underline{\Omega}^\grast_\der(\algA) = \Omega^\grast_\der(\algA) = \Omega^\grast(\varM) \otimes \Omega^\grast_\der(M_n)$ with the differential $\hd = \dd + \dd'$, where $\dd$ is the de~Rham differential and $\dd'$ is the differential introduced in the previous example.

\item The noncommutative $1$-form $i\theta$ is defined as $i\theta(X + \adrep_\gamma) = \gamma$. It splits the short exact sequence of Lie algebras and $C^\infty(\varM)$-modules
\begin{equation}
\label{eq-splittingsecderivationstrivialcase}
\xymatrix@1@C=25pt{{\algzero} \ar[r] & {\algA_0} \ar[r] & {\der(\algA)} \ar[r] \ar@/_0.7pc/[l]_-{i\theta}& {\Gamma(\varM)} \ar[r] & {\algzero}}
\end{equation}

\item Noncommutative integration is a well-defined map of differential complexes
\begin{align*}
\int_{\text{n.c.}} : \Omega^\grast_\der(\algA) & \rightarrow \Omega^{\grast - (n^2-1)}(\varM)
&
\int_{\text{n.c.}} \hd \omega &= \dd \int_{\text{n.c.}} \omega
\end{align*}

\end{itemize}
\end{proposition} 

Using a metric $h$ on $\varM$ and the metric $g_{k \ell} = \frac{1}{n} \tr(E_k E_\ell)$ on the matrix part, one can define a metric on $\der(\algA)$ as follows,
\begin{equation*}
\widehat{g}(X+ \adrep_\gamma, Y + \adrep_\eta) = h(X,Y) + \textstyle \frac{1}{m^2}g(\gamma \eta)
\end{equation*}
where $m$ is a positive constant which measures the relative ``weight'' of the two ``spaces''. In physical natural units, it has the dimension of a mass.

\bigskip
Consider now the right $\algA$-module $\modM = \algA$. As for the algebra $M_n$, the noncommutative $1$-form $-i\theta$ defines a canonical noncommutative connection by the relation $\widehat{\nabla}^{-i\theta}_\kX a = \kX a - i\theta(\kX) a$ for any $a \in \algA$.

\begin{proposition}[Properties of $\widehat{\nabla}^{-i\theta}$]
For any $a \in \algA$ and $\kX = X + \adrep_\gamma \in \der(\algA)$, one has $\widehat{\nabla}^{-i\theta}_\kX a = X \cdotaction a - a \gamma$.

The curvature of the noncommutative connection $\widehat{\nabla}^{-i\theta}$ is zero.

The gauge transformed connection $\widehat{\nabla}^{-i\theta g}$ by $g \in C^\infty(\varM) \otimes GL_n(\gC)$ is associated to the noncommutative $1$-form $\kX \mapsto -i\theta(\kX) + g^{-1} (X\cdotaction g) = - \gamma + g^{-1} (X\cdotaction g)$.
\end{proposition}

\section{The endomorphism algebra of a vector bundle}

The second example of the previous section mixes together two geometries: the de~Rham ordinary differential geometry, and the noncommutative derivation-based differential geometry of the matrix algebra. This last geometry is very similar to the ordinary geometry of the Lie group $SL_n(\gC)$.

It is common in physics to consider the geometry of a based manifold with the geometry of a Lie group (especially Lie groups of the type $SU(n)$): indeed, this is the geometry underlying gauge theories as they are used in the Standard Model of particle physics. This kind of geometry is well understood in the context of principal fiber bundles (see Section~\ref {Abriefreviewofordinaryfiberbundletheory}).

This section is devoted to the definition of a noncommutative geometry which generalizes and contains in a precise meaning (see Section~\ref{Relationswiththeprincipalfiberbundle}) some essential aspects of the ordinary geometry of $SU(n)$-principal fiber bundles.

\subsection{The algebra and its derivations}

Let $\varE$ be a $SU(n)$-vector bundle over $\varM$ with fiber $\gC^n$. Consider $\End(\varE)$, the fiber bundle of endomorphisms of $\varE$ (see Example~\ref{ex-Theendomorphismbundle}). We denote by $\algA$ the algebra of sections of $\End(\varE)$. This is the algebra we will study using noncommutative differential geometry. 

For later references, the trivial case is the situation where $\varE = \varM \times \gC^n$ is the trivial fiber bundle. In that case, one has $\algA = C^\infty(\varM)\otimes M_n$. Its noncommutative geometry is the one exposed as the second example of the previous section. In general, $\algA$ is (globally) more complicated.

Let us motivate the importance of this algebra by the following remarks:
\begin{remark}[Relation to ordinary geometry]
The endomorphism fiber bundle $\End(\varE)$ is associated to a $SU(n)$-principal fiber bundle $\varP$ for the couple $(M_n, \Ad)$.

Because $G=SU(n) \subset M_n(\gC)$ and $\kg = \ksu(n) \subset M_n(\gC)$, one has $\varP \times_\alpha G \subset \End(\varE)$ and $\Ad \varP = \varP \times_\Ad \kg  \subset \End(\varE)$ where $\alpha_g(h) = g^{-1} h g$ for any $g,h \in G$.

This implies that the gauge group $\caG = \Gamma(\varP \times_\alpha G)$ and its Lie algebra $\lie\caG = \Gamma(\Ad \varP)$ (see Example~\ref{ex-ThegaugegroupanditsLiealgebra}) are subspaces of $\algA$.

We will see in the following that (ordinary) connections are also related to this noncommutative geometry.
\end{remark} 

Locally, using trivialisations of $\varE$, the algebra $\algA$ looks like $C^\infty(U) \otimes M_n$. This is very useful to study some objects defined on $\algA$.

\begin{proposition}[Basic properties]
One has $\caZ(\algA) = C^\infty(\varM)$.

Involution, trace map and determinant ($\tr, \det : \algA \rightarrow C^\infty(\varM)$), are well defined fiberwise. 

Let us define $SU(\algA)$ as the unitaries in $\algA$ of determinant $1$, and $\ksu(\algA)$ as the traceless antihermitean elements. Then $\caG = SU(\algA)$ and $\lie\caG = \ksu(\algA)$.
\end{proposition} 

This identifies exactly the gauge group and its Lie algebra as natural and canonical subspaces of $\algA$.

\bigskip
Let $\rho : \der(\algA) \rightarrow \der(\algA)/\Int(\algA) = \Out(\algA)$ be the projection of the short exact sequence \eqref{eq-secderivations}. This projection has an natural interpretation in this context:
\begin{proposition}[The derivations of $\algA$]
\label{prop-thederivationsodalgA}
One has $\Out(\algA) \simeq \der(C^\infty(\varM)) = \Gamma(\varM)$ and $\rho$ is the restriction of derivations $\kX \in \der(\algA)$ to $\caZ(\algA) = C^\infty(\varM)$. $\Int(\algA)$ is isomorphic to $\algA_0$, the traceless elements in $\algA$.

The short exact sequence of Lie algebras and $C^\infty(\varM)$-modules of derivations looks like
\begin{equation*}
\xymatrix@R=0pt@C=15pt{ 
{\algzero} \ar[r] & {\Int(\algA)} \ar[r] & {\der(\algA)} \ar[r]^-{\rho} & {\Gamma(\varM)} \ar[r] & {\algzero}\\
           &                    &  \kX        \ar@{|->}[r]   & X       &
}
\end{equation*}
Real inner derivations are given by the $\adrep_\xi$ with $\xi \in \lie\caG = \ksu(\algA)$.
\end{proposition} 

The short exact sequence in this proposition describes the general situation which generalises the splitting for the trivial situation encountered in \eqref{eq-splittingsecderivationstrivialcase}. There is no \textsl{a priori} canonical splitting in the non trivial case. Moreover, the noncommutative $1$-form $i\theta$ is no more defined here. But one can define a map of $C^\infty(\varM)$-modules:
\begin{align*}
i\theta : \Int(\algA) &\rightarrow \algA_0
&
\adrep_\gamma &\mapsto \textstyle\gamma - \frac{1}{n}\tr(\gamma) \bbbone
\end{align*}

Here is an important result which can be proved using local trivialisations:
\begin{proposition}
\begin{equation*}
\underline{\Omega}^\grast_\der(\algA) = \Omega^\grast_\der(\algA)
\end{equation*}
\end{proposition}

The next proposition will be used in the study of ordinary connections on $\varE$ and their relations to the noncommutative geometry of $\algA$:
\begin{Proposition}[Horizontal forms for the operation of $\Int(\algA)$]
\label{prop-HorizontalformsfortheoperationofInt(algA)}
The space of sections $\Gamma(\exter^\grast T^\ast \varM \otimes \End(\varE))$ is the graded algebra of noncommutative horizontal forms in $\Omega^\grast_\der(\algA)$ for the operation of $\Int(\algA)$ on $\Omega^\grast_\der(\algA)$.
\end{Proposition}

\subsection{Ordinary connections}

Let us now show how this noncommutative geometry is well adapted to not only study ordinary connections on the vector bundle $\varE$, but also,as will be seen in the next section, to allow to some natural generalisations of these connections.

Let $\nabla^\varE$ be any (usual) connection on the vector bundle $\varE$. One can define the two associated connections $\nabla^{\varE^\ast}$ on $\varE^\ast$ and $\nabla$ on $\End(\varE)$ by the relations
\begin{align*}
X\cdotaction \langle \varphi, s \rangle &= \langle \nabla_X^{\varE^\ast} \varphi,  s \rangle + \langle \varphi, \nabla_X^\varE s \rangle
&
\nabla_X (\varphi \otimes s) &= (\nabla_X^{\varE^\ast} \varphi) \otimes s + \varphi \otimes (\nabla_X^\varE s)
\end{align*}
with $X \in \Gamma(\varM)$, $\varphi \in \Gamma(\varE^\ast)$ and $s \in \Gamma(\varE)$

In the following, we will use the notation $X = \rho(\kX) \in \Gamma(\varM)$ for any $\kX \in \der(\algA)$.

\begin{proposition}[The noncommutative $1$-form $\alpha$]
\label{prop-Thenoncommutative1formalpha}
For any $X \in \Gamma(\varM)$, $\nabla_X$ is a derivation of $\algA$. 

For any $\kX \in \der(\algA)$, the difference $\kX - \nabla_X$ is an inner derivation. This permits one to introduce $\kX \mapsto \alpha(\kX)= -i\theta(\kX - \nabla_X)$. By construction, $\alpha$ is a noncommutative $1$-form $\alpha \in\Omega^1_\der(\algA)$ which gives the decomposition
\begin{equation*}
\kX = \nabla_X - \adrep_{\alpha(\kX)}
\end{equation*}

For any $\gamma \in \algA_0$, one has $\alpha(\adrep_\gamma) = -\gamma$, for any $\kX \in \der(\algA)$, one has $\tr \alpha(\kX)= 0$, and for any $\kX \in \der_\gR(\algA)$ one has $\alpha(\kX)^\ast + \alpha(\kX) = 0$.
\end{proposition} 

Notice that by the decomposition given in this proposition, $X\mapsto \nabla_X$ is a splitting as $C^\infty(\varM)$-modules of the short exact sequence
\begin{equation}
\label{eq-splittingofshortexactsequenceofderivations}
\xymatrix@1@C=25pt{{\algzero} \ar[r] & {\algA_0} \ar[r] & {\der(\algA)} \ar[r] & {\Gamma(\varM)} \ar[r] \ar@/_0.7pc/[l]_-{\nabla} & {\algzero}}
\end{equation}
The obstruction to be a splitting of Lie algebras is nothing but the curvature of $\nabla$ which we denote by $R(X,Y) = [\nabla_X, \nabla_Y] - \nabla_{[X,Y]}$.

\begin{remark}[$\alpha$ extends $-i \theta$]
The relation $\alpha(\adrep_\gamma) = -\gamma$ shows that $\alpha$ extends $-i \theta :  \Int(\algA) \rightarrow \algA_0$. As will be seen in Proposition~\ref{prop-ordinaryconnectionsandnoncommutativeforms}, any such extension is indeed related to a choice of an ordinary connection of $\varE$.
\end{remark}

One can then introduce the main result which connects the ordinary geometry of $\varE$ and the noncommutative differential geometry of $\algA$:
\begin{proposition}[Ordinary connections and noncommutative forms]
\label{prop-ordinaryconnectionsandnoncommutativeforms}
The map $\nabla^\varE \mapsto \alpha$ is an isomorphism between the affine spaces of $SU(n)$-connections on $\varE$ and the traceless antihermitean noncommutative $1$-forms on $\algA$ such that $\alpha(\adrep_\gamma) = -\gamma$.

The noncommutative $2$-form $(\kX, \kY) \mapsto \Omega(\kX, \kY) = \hd\alpha(\kX, \kY) + [\alpha(\kX), \alpha(\kY) ]$ depends only on the projections $X$ and $Y$ of $\kX$ and $\kY$. This means that it is a horizontal noncommutative $2$-form for the operation of $\Int(\algA)$ on $\Omega^\grast_\der(\algA)$.

The curvature $R^\varE$ of $\nabla^\varE$, considered as a section of $\exter^2 T^\ast \varM \otimes \Ad \varP \subset \exter^2 T^\ast \varM \otimes \End(\varE)$ (see Proposition~\ref{prop-HorizontalformsfortheoperationofInt(algA)}), is exactly the horizontal noncommutative $2$-form $\Omega$.
\end{proposition} 

\begin{remark}[The intermediate construction in ordinary geometry]
\label{rmk-Theintermediateconstructioninordinarygeometry}
We saw in Remark~\ref{rmk-intermediateconstruction} that in the ordinary geometry of a principal fiber bundle, one is used to introduce connections as $1$-forms $\omega \in \Omega^1(\varP) \otimes \kg$, with two conditions: vertical normalisation and equivariance. Its curvature is then a $2$-form in $\Omega^2(\varP) \otimes \kg$, equivariant and horizontal. The other possibility is to introduce a family of local $1$-forms $A \in \Omega^1(U) \otimes \kg$ on open subsets $U$ of trivialisations of $\varP$, with some non homogeneous gluing relations. The curvature is represented by a family of $2$-forms $F \in \Omega^2(U)\otimes \kg$ satisfying some homogeneous gluing relations. 

Using the ``top'' construction (equivariant and horizontal properties) or the ``bottom'' one (homogeneous gluing relations), one can show that the curvature is indeed a section of the vector bundle $\exter^2 T^\ast \varM \otimes \Ad \varP \subset \exter^2 T^\ast \varM \otimes \End(\varE)$.

This proposition shows that this ``intermediate'' construction (the curvature as a section of a vector bundle) can be completed at the level of the connection $1$-form, at the price of using noncommutative geometry (the noncommutative $1$-form $\alpha$) in order to take into account the non homogeneous gluing relations of the local connection $1$-forms (see Remark~\ref{remark-localexpressions}). The vertical normalisation and the equivariant conditions at the level of $\varP$ are replaced by a unique condition on inner derivations at the level of $\algA$.
\end{remark} 

Let us now look at gauge transformations. Let $u \in \caG = SU(\algA)$ and $\xi \in \lie\caG = \ksu(\algA)$. 
\begin{proposition}[Gauge transformations]
The noncommutative $1$-form $\alpha^u$ corresponding to the gauge transformed connection $\nabla^{\varE u}$ is given by the suggestive expression
\begin{equation*}
\alpha^u = u^\ast \alpha u + u^\ast \hd u
\end{equation*}

The infinitesimal gauge transformation induced by $\xi$ is
\begin{equation*}
\alpha \mapsto -\hd \xi - [\alpha, \xi] = L_{\adrep_\xi} \alpha
\end{equation*}
This means that we can interpret infinitesimal gauge transformations on connections on $\varE$ as Lie derivative of real inner derivations on $\algA$.
\end{proposition} 

\bigskip
\begin{Remark}[Local expressions]
\label{remark-localexpressions}
It is instructive to look at the noncommutative $1$-form $\alpha$ in some local trivialisation of $\varE$.  Let $U_i \subset \varM$ be a local trivialisation system of $\varE$, and so of $\End(\varE)$. We denote by $a^\loc_i : U_i \rightarrow M_n$ the restriction of the global section $a \in \algA$ looked at in a local trivialisation.  

Over $U_i \cap U_j \neq \ensvide$, one has the homogeneous gluing relations $a^\loc_j = \Ad_{g_{ij}^{-1}} a^\loc_i = g_{ij}^{-1} a^\loc_i g_{ij}$, with $g_{ij} : U_i \cap U_j \rightarrow SU(n)$ the transition functions.

Locally a derivation $\kX \in \der(\algA)$ can be written as $\kX^\loc_i = X_i + \adrep_{\gamma_i}$, with $\gamma_i : U_i \rightarrow M_n$ (traceless) and $X_i$ a vector fields on $U_i$. Using the map $\rho$, one gets that $X_i$ is the restriction of $X=\rho(\kX)$ to $U_i$, so that we can write $X = X_i$.

Using compatibility with the homogeneous gluing relations for sections, one finds that the $\gamma_i$'s satisfy some non homogeneous gluing relations
\begin{equation*}
\gamma_j = g^{-1}_{ij} \gamma_i g_{ij} + g^{-1}_{ij} X\cdotaction g_{ij}
\end{equation*}
The noncommutative $1$-form $\alpha$ is then locally given by the expressions
\begin{equation*}
\alpha^\loc_i(X + \adrep_{\gamma_i}) = A_i(X) - \gamma_i
\end{equation*}
where the $A_i$'s form the family of local trivialisation of the connection $1$-form. It is then easy to check that 
\begin{align*}
\alpha^\loc_j(X + \adrep_{\gamma_j}) &= A_j(X) - \gamma_j = (g^{-1}_{ij} A_i(X) g_{ij} + g^{-1}_{ij} X \cdotaction g_{ij}) - (g^{-1}_{ij} \gamma_i g_{ij} + g^{-1}_{ij} X\cdotaction g_{ij}) \\
&= g^{-1}_{ij} (A_i(X) - \gamma_i) g_{ij} = g^{-1}_{ij} \alpha^\loc_i(X + \adrep_{\gamma_i}) g_{ij}
\end{align*}
so that these expressions indeed define a global section in $\algA$.

As can be noticed here, the global existence of the noncommutative $1$-form $\alpha$ relies on the fact that the $A_i$'s and the $\gamma_i$'s share the same non homogeneous gluing relations.
\end{Remark}

\section{\texorpdfstring{Noncommutative connections on $\algA$}{Noncommutative connections on A}}

In this section, we mainly study noncommutative connections on the right $\algA$-module $\modM=\algA$ equipped with the canonical Hermitean structure $(a,b) \mapsto a^\ast b$. 

\subsection{Main properties}

As we saw in Proposition~\ref{prop-NoncommutativeconnectionsonmodM=algA}, a noncommutative connection $\widehat{\nabla}$ on the right $\algA$-module $\modM=\algA$ is given by a noncommutative $1$-form $\omega \in \Omega^1_\der(\algA)$ by the relation $\widehat{\nabla}_\kX a = \kX a + \omega(\kX) a$. This implies that studying $\widehat{\nabla}$ is equivalent to studying $\omega$.

Let us first look at some particular noncommutative connections:
\begin{proposition}[The noncommutative connection associated to $\alpha$]
Let $\nabla^\varE$ be a $SU(n)$-connection on $\varE$, and denote by $\alpha$ its associated noncommutative $1$-form. Then, the noncommutative connection $\widehat{\nabla}^\alpha$ defined by the noncommutative $1$-form $\alpha$ is given by
\begin{equation}
\label{eq-expressionofwidehatnablaalpha}
\widehat{\nabla}^\alpha_\kX a = \nabla_X a + a \alpha(\kX)
\end{equation}
In particular, for any $X \in \Gamma(\varM)$, one has $\widehat{\nabla}^\alpha_{\nabla_X} a = \nabla_X a$.

This noncommutative connection $\widehat{\nabla}^\alpha$ is compatible with the canonical Hermitean structure.

The curvature of $\widehat{\nabla}^\alpha$ is $\hR^\alpha(\kX, \kY) = R^\varE(X,Y)$.

A gauge transformation induced by $u \in \caG = SU(\algA)$ on the connection $\nabla^\varE$ induces a (noncommutative) gauge transformation on $\widehat{\nabla}^\alpha$.
\end{proposition} 

\begin{proof}
Recall that by definition, one has $\kX = \nabla_X - \adrep_{\alpha(\kX)}$ and $\widehat{\nabla}^\alpha_\kX a = \kX a + \alpha(\kX) a$. This proves \eqref{eq-expressionofwidehatnablaalpha}.

One the other hand, the curvature of $\widehat{\nabla}^\alpha$ is the noncommutative $2$-form $\hd\alpha (\kX, \kY) + [ \alpha(\kX), \alpha(\kY) ]$ which has been identified with the curvature of $\nabla$ in Proposition~\ref{prop-ordinaryconnectionsandnoncommutativeforms}.

In a gauge transformation, one has $\alpha^u = u^\ast \alpha u + u^\ast \hd u$, which is also the noncommutative gauge transformation applied to $\widehat{\nabla}^\alpha$.
\end{proof}

We now arrive at the main result of this report:

\begin{theorem}[Ordinary connections as noncommutative connections]
The space of noncommutative connections on the right $\algA$-module $\algA$ compatible with the Hermitean structure $(a,b) \mapsto a^\ast b$ contains the space of ordinary $SU(n)$-connections on $\varE$. 

This inclusion is compatible with the corresponding definitions of curvature and gauge transformations.
\end{theorem} 

From now on, one can consider that an ordinary connection is a noncommutative connection on the right $\algA$-module $\algA$. In this respect, this point of view generalizes the notion of connection through the intermediate construction.

A natural question is: what are noncommutative connections from a physical point of view?

\subsection{Decomposition of noncommutative connections on the module $\algA$}

In order to answer the above question, one can look at some natural decompositions of noncommutative connections, and compare these decompositions to ``ordinary'' connections.

Let us fix a connection $\nabla^\varE$ on $\varE$, and denote by $\alpha$ its associated noncommutative $1$-form. Then any noncommutative connection $\widehat{\nabla}$ can be decomposed as
\begin{equation*}
\widehat{\nabla}_\kX a = \widehat{\nabla}^\alpha_\kX a + \caA(\kX) a
\end{equation*}
with $\caA \in \Omega^1_\der(\algA)$, so that $\omega = \alpha + \caA$ is the noncommutative $1$-form for $\widehat{\nabla}$.

Using the relation $\kX = \nabla_X - \adrep_{\alpha(\kX)}$, one splits $\caA$ as $\caA(\kX) = \ka(X) - \kb(\alpha(\kX))$, where $\kb : \algA_0 \rightarrow \algA$ is defined by $\kb(\gamma) = \caA(\adrep_\gamma)$.

A straightforward computation shows that the curvature of $\widehat{\nabla}$ can then be written as
\begin{align*}
\hR(\kX, \kY) &= R^\varE(X,Y) + \nabla_X \caA(\kY) - \nabla_Y \caA(\kX) - \caA([\kX, \kY]) + [ \caA(\kX), \caA(\kY)]\\
&= R^{\varE, \ka}(X,Y) - \nabla^\ka_X \kb (\alpha(\kY)) + \nabla^\ka_Y \kb (\alpha(\kX))\\
&\quad\quad\quad\quad\quad\quad\quad\quad\quad + [\kb (\alpha(\kX)), \kb (\alpha(\kY))] + \kb (\alpha([\kX, \kY]))
\end{align*}
where $R^{\varE, \ka}$ is the curvature of the connection $\nabla^{\varE, \ka}_X s = \nabla^\varE_X s + \ka(X) s$ on $\varE$ and $\nabla^\ka$ is its associated connection on $\End(\varE)$.

Performing a gauge transformation with $u \in \caG = SU(\algA)$, one has
\begin{align*}
\caA^u &= u^\ast \caA u + u^\ast (\nabla u)
&
\ka^u &= u^\ast \ka u + u^\ast (\nabla u)
&
\kb^u &= u^\ast \kb u
\end{align*}

Notice the replacement of the differential by $\nabla$ in these expressions.

\begin{remark}[Local expressions]
In Remark~\ref{remark-localexpressions}, we looked at local expressions of the noncommutative $1$-form $\alpha$. Let us now look at the previous decomposition in a local trivialisation of $\varE$. The noncommutative connection $\widehat{\nabla}$ takes the local expression:
\begin{equation*}
\widehat{\nabla}^\loc_{\kX^\loc} a^\loc = X \cdotaction a^\loc + \left[ A(X) + \ka^\loc(X) - \kb^\loc(A(X)) \right] a^\loc + \kb^\loc(\gamma) a^\loc - a^\loc \gamma
\end{equation*}
where $A$ is the local connection $1$-form of $\nabla^\varE$ and $\kX^\loc = X + \adrep_\gamma$ as before.

In a change of local trivialisation, the two local maps $\gamma \mapsto \kb^\loc(\gamma)$ and $X \mapsto \ka^\loc(X)$ transform as
\begin{align*}
\gamma &\mapsto \kb'^\loc(\gamma) = g^{-1} \kb^\loc(g \gamma g^{-1}) g
&
X &\mapsto \ka'^\loc(X) = g^{-1} \ka^\loc(X) g
\end{align*}
which are both homogeneous gluing relations.
\end{remark}

\bigskip
In order to be more explicit, consider now the trivial situation $\varE = \varM \times \gC^n$ and $\algA = C^\infty(\varM) \otimes M_n$.

As a reference (ordinary) connection, one can take $\nabla^\varE_X s= X\cdotaction s$, so that, using the local expression of $\alpha$, one has 
\begin{equation*}
\alpha(\kX) = \alpha(X + \adrep_\gamma) = - \gamma = -i \theta (\kX)
\end{equation*}
with $\tr \gamma = 0$. Then $\widehat{\nabla}^\alpha = \widehat{\nabla}^{-i\theta}$. Moreover, $\kb(\alpha(\kX)) = \kb(- \gamma) = - \kb(\gamma)$, so that
\begin{equation*}
\widehat{\nabla}_\kX a = X \cdotaction a + \ka(X) a + \kb(\gamma) a = \overline{\nabla}^\ka_X a + \kb(\gamma) a
\end{equation*}
where $\overline{\nabla}^\ka$, defined in some local trivialization by $\overline{\nabla}^\ka_X a = X \cdotaction a + \ka(X) a$, is an ordinary connection on $\End(\varE)$, but is not $\nabla^\ka$, which takes the explicit local form $\nabla^\ka_X a = X \cdotaction a + [\ka(X), a]$. 

$X \mapsto \ka(X)$ behaves like a gauge potential with respect to gauge transformations (here $\nabla = \dd$). The difference between ordinary connections and noncommutative connections is the presence of $\kb$, which represents some additional fields in physics. These fields have homogeneous gauge transformations.

The curvature can be written, for $\kX = X + \adrep_\gamma$ and $\kY = Y + \adrep_\eta$,
\begin{equation*}
\hR(\kX, \kY) = R^{\varE, \ka}(X,Y) + (\widetilde{\nabla}^\ka_X \kb)(\eta) - (\widetilde{\nabla}^\ka_Y \kb)(\gamma) + [\kb(\gamma), \kb(\eta)] - \kb([\gamma, \eta])
\end{equation*}
where $\widetilde{\nabla}^\ka$ is the connection $(\widetilde{\nabla}^\ka_X \kb)(\eta) = X\cdotaction \kb(\eta) - \kb( X \cdotaction \eta) + [\ka(X), \kb(\eta)]$ on the space of $C^\infty(\varM)$-linear maps $\algA_0 \rightarrow \algA$.

\subsection{Yang-Mills-Higgs Lagrangian on the module $\algA$}

Consider, as before, the trivial case $\algA = C^\infty(\varM) \otimes M_n$ and the right $\algA$-module $\algA$. Let $\ka = \ka_\mu \dd x^\mu$ and $\kb = \kb_k \theta^k$, with $\ka_\mu, \kb_k \in C^\infty(\varM) \otimes M_n$.

The curvature is then the noncommutative $2$-form
\begin{equation*}
\hR = \textstyle\frac{1}{2}(\partial_\mu \ka_\nu - \partial_\nu \ka_\mu + [ \ka_\mu, \ka_\nu])\dd x^\mu \dd x^\nu + (\partial_\mu \kb_k + [ \ka_\mu, \kb_k]) \dd x^\mu \theta^k + \frac{1}{2}([\kb_k, \kb_\ell] - C^m_{k \ell} \kb_m) \theta^k \theta^\ell
\end{equation*}

Using a metric (here euclidean) on $\der(\algA)$ and an associated Hodge star operation, one can define a Lagrangian. Using ordinary and noncommutative integration, one then defines the action:
\begin{multline*}
S(\hR) = \int \dd x \tr\bigg\{ \sum_{\mu,\nu}{\textstyle\frac{1}{4}}(\partial_\mu \ka_\nu - \partial_\nu \ka_\mu + [ \ka_\mu, \ka_\nu])^2 \\[-10pt]
+ m^2 \sum_{\mu,k} (\partial_\mu \kb_k + [ \ka_\mu, \kb_k])^2 + m^4 \sum_{k, \ell}{\textstyle\frac{1}{4}} ([\kb_k, \kb_\ell] - C^m_{k \ell} \kb_m)^2 \bigg\}
\end{multline*}

The integrand is zero when
\begin{align*}
\ka &\text{ gauge equivalent to $0$}
&
\dd \kb &= 0
&
[\kb_k, \kb_\ell] &= C^m_{k \ell} \kb_m
\end{align*}
so that the $\kb_k$'s are constant and induce a representation of $\ksl_n$ in $M_n$.

For the right $\algA$-module $\modM = C^\infty(\varM) \otimes M_{r,n}$, one would get similar results: flat connections are classified by inequivalent representations of $\ksl_n$ in $M_r$.

\begin{Remark}[Physical interpretation]
From a fields theory point of view, one can notice that the $\ka_\mu$ fields behave like ordinary Yang-Mills fields, for a $SU(n)$ gauge theory. On the other hand, the interesting point is that the $\kb_k$ fields behave as Higgs fields: in the above action, the vacuum states can be non trivial and the Higgs mechanism of mass generation is possible. Finally, the coupling between these fields is a covariant derivative in the adjoint representation.
\end{Remark}

For a more general situation where $\algA$ is not the trivial case, one can proceed in the same line:
\begin{itemize}
\item One has to use a reference connection on $\varE$ to help to decompose noncommutative connections.
\item The curvature looks similar except for the presence of the reference connection.
\item The Hodge star operator is defined.
\item The action splits into three terms, and the vacuum states are related to the global structure of the vector fiber bundle $\varE$.
\end{itemize}





\section{Relations with the principal fiber bundle}
\label{Relationswiththeprincipalfiberbundle}

It is possible to look at the noncommutative geometry of $\algA$ using the ordinary geometry of the underlying $SU(n)$-principal fiber bundle $\varP$ and the noncommutative geometry of a bigger algebra, hereafter denoted by $\algB$.

\subsection{The algebra $\algB$}

As before, let $\varP$ be the $SU(n)$-principal fiber bundle to which $\varE$ is associated, and consider the associative algebra $\algB = C^\infty(\varP) \otimes M_n$. This algebra is an example of the trivial situation mentioned in \ref{subsubsection-ThealgebraalgA=Cinfty(varM)otimesMn}, so that one has immediately the following facts: the center of $\algB$ is $\caZ(\algB) = C^\infty(\varP)$, its Lie algebra and $\caZ(\algB)$-module of derivations splits, $\der(\algB) = \Gamma(\varP) \oplus [C^\infty(\varP) \otimes \ksl_n]$, and its noncommutative differential calculus is the tensor product of the two differential calculi associated to $\varP$ and $M_n$: $\Omega^\grast_\der(\algB) = \Omega^\grast(\varP) \otimes \Omega^\grast_\der(M_n)$ with the differential $\hd = \dd + \dd'$.

One can embed the real Lie algebra $\ksu(n)$ as a subalgebra of $\der(\algB)$ in two ways:
\begin{align*}
\xi &\mapsto \xi^v \text{ vertical vector field on $\varP$}
&
\xi &\mapsto \adrep_\xi \text{ inner derivation}
\end{align*}
This permits one to introduce the following two Lie subalgebras of $\der(\algB)$: 
\begin{align*}
\kg_\adrep &= \{ \adrep_\xi \ / \ \xi \in \ksu(n) \} &
\kg_\equ &= \{ \xi^v + \adrep_\xi \ / \ \xi \in \ksu(n) \}
\end{align*}

\begin{proposition}
The algebra $C^\infty(\varP)$ (resp. $\algA$) is the set of invariant elements for the action of $\kg_\adrep$ (resp. $\kg_\equ$) on $\algB$.
\end{proposition}

\begin{proof}
$C^\infty(\varP)$ is the invariants of $\kg_\adrep$ because $\adrep_\xi b = 0$ for any $\xi \in \ksu(n)$ implies $b \in \caZ(\algB)$. $\algA$ is the invariants of $\kg_\equ$ because $\algA$ is the set of sections of $\End(\varE)$, which is $\caF_{SU(n)}(\varP,M_n)$, the space of $SU(n)$-equivariant maps from $\varP$ to $M_n$. The relation $\xi^v \cdotaction b + \adrep_\xi b = 0$ for any $\xi \in \ksu(n)$ is the infinitesimal version of this equivariance.
\end{proof}

The two Lie subalgebras $\kg_\adrep$ and $\kg_\equ$ define Cartan operations on $(\Omega^\grast_\der(\algB), \hd)$. The previous proposition tells us that the algebras $\algB$, $C^\infty(\varP)$ and $\algA$ are related by these two operations. 

Moreover, $C^\infty(\varM)$ is itself the set of invariant elements for $\xi \mapsto \xi^v$ in $C^\infty(\varP)$ and the invariants in $\algA$ for the operation of $\Int(\algA)$.

\begin{proposition}[Relations between the differential calculi]
At the level of differential calculi, all these relations generalize in the following structure:
\begin{equation*} 
\xymatrix@R=30pt@C=70pt@M=5pt{
{\Omega^\grast(\varP) \otimes \Omega^\grast_\der(M_n)}  
&  {\Omega^\grast(\varP)}
  \ar@{_{(}->}[l]_-{\text{basic elements}}^-{\text{$\ksu(n) \ni \xi \mapsto \adrep_\xi$}} 
\\ 
{\Omega^\grast_\der(\algA)}
  \ar@{^{(}->}[u]^-{\substack{\text{basic elements} \\ \text{$\ksu(n) \ni \xi \mapsto \xi^v + \adrep_\xi$}}}
&  {\Omega^\grast(\varM)}
  \ar@{^{(}->}[u]_-{\substack{\text{basic elements} \\ \text{$\ksu(n) \ni \xi \mapsto \xi^v$}}} 
  \ar@{_{(}->}[l]_-{\text{basic elements}}^-{\Int(\algA)} }
\end{equation*} 
\end{proposition}

In order to show these relations, one need the concept of noncommutative quotient manifold introduced in \cite{Mass:07}. We refer to \cite{Mass:15} for the complete proof.

Notice that this proposition contains a well known result in ordinary differential geometry, which says that the space of tensorial forms in $\Omega^\grast(\varP) \otimes \kg$ (horizontal and equivariant for the action induced by right multiplication on $\varP$ and the adjoint action on the Lie algebra $\kg$) is the space $\Omega^\grast(\varM, \Ad \varP)$ of forms on the base manifold $\varM$ with values in the vector bundle $\Ad \varP$. This result permits one to indentify the curvature of a connection on $\varP$ to a form in $\Omega^2(\varM, \Ad \varP)$ (see Proposition~\ref{prop-ordinaryconnectionsandnoncommutativeforms} and Remark~\ref{rmk-Theintermediateconstructioninordinarygeometry}).

In Proposition~\ref{prop-Generalpropertiesofthedifferentialcalculustrivialcase}, we saw that the noncommutative integration is well defined on algebras like $\algB$. This induces a map
\begin{equation*}
\int_{\text{n.c.}} : \Omega^{r}_\der(\algB) \rightarrow \Omega^{r-(n^2-1)}(\varP)
\end{equation*}
which has the following properties:
\begin{proposition}[Noncommutative integration]
If $\omega \in \Omega^{r}_\der(\algB)$ is a horizontal (resp. basic) noncommutative form for one of the operations of $\kg_\adrep$ or $\kg_\equ$, then $\int_{\text{n.c.}} \omega \in \Omega^{r-(n^2-1)}(\varP)$ is horizontal (resp. basic) for the corresponding operation restricted to $\Omega^{\grast}(\varP) \subset \Omega^{\grast}_\der(\algB)$.

This noncommutative integration then restricts to a ``noncommutative integration along the noncommutative fiber'' $\Omega^{\grast}_\der(\algA) \rightarrow \Omega^{\grast-(n^2-1)}(\varM)$.

This noncommutative integration is compatible with the differentials, and it induces maps in cohomologies
\begin{align*}
\int_{\text{n.c.}} : H^\grast(\Omega^{\grast}_\der(\algB), \hd) & \rightarrow H^{\grast-(n^2-1)}_{\text{dR}}(\varP)
\\
\int_{\text{n.c.}} : H^\grast(\Omega^{\grast}_\der(\algA), \hd) & \rightarrow H^{\grast-(n^2-1)}_{\text{dR}}(\varM)
\end{align*}

\end{proposition} 

This situation looks very similar to the integration along the fibers of compactly supported (along the fibers) differential forms in the theory of vector bundles.

\subsection{\texorpdfstring{Ordinary {\itshape vs.} noncommutative connections}{Ordinary vs. noncommutative connections}}

It is instructive to identify ordinary connections in this setting. Let $\widehat{\nabla}$ be a noncommutative connection on the right $\algA$-module $\algA$, and denote by $\alpha \in \Omega^1_\der(\algA)$ it associated noncommutative $1$-form.

As a basic noncommutative $1$-form in $\Omega^1_\der(\algB)$ for the operation of $\kg_\equ$, one can write
\begin{equation*}
\alpha = \omega - \phi \in [\Omega^1(\varP) \otimes M_n] \oplus [C^\infty(\varP) \otimes M_n \otimes \ksl_n^\ast]
\end{equation*}
The basic condition implies the relations
\begin{align*}
(L_{\xi^v} + L_{\adrep_\xi}) \omega &= 0 
&
(L_{\xi^v} + L_{\adrep_\xi}) \phi &= 0 
&
i_{\xi^v} \omega - i_{\adrep_\xi} \phi &= 0
\end{align*}
for any $\xi \in \ksu(n)$.

\begin{proposition}[Ordinary connection]
Let $\nabla^\varE$ be an ordinary connection on $\varE$ and $\alpha \in \Omega^1_\der(\algA)$ its associated noncommutative $1$-form. Then, as a basic element in $\Omega^1_\der(\algB)$, one has
\begin{equation*}
\alpha = \omega - i\theta
\end{equation*}
where $\omega \in \Omega^1(\varP) \otimes \ksu(n) \subset \Omega^1(\varP) \otimes M_n$ is the connection $1$-form on $\varP$ associated to $\nabla^\varE$ and $i\theta$ is the canonical noncommutative $1$-form defined in $\Omega^1_\der(\algB)$ (Proposition~\ref{prop-Generalpropertiesofthedifferentialcalculustrivialcase}).
\end{proposition} 

In order to prove this formula, one has to use the equivariance and the vertical condition for $\omega$, and some of the properties listed before on $i\theta$.

Notice that this inclusion of ordinary connection into the space of basic $1$-forms on $\algB$ is canonical, since the noncommutative $1$-form $i\theta$ is itself canonical.

\subsection{Splittings coming from connections}

The previous considerations show how the differential calculi connect together through some Cartan operations. There also exist some strong relations between the derivations of $\algA$, some derivations of $\algB$, and some vector fields on $\varP$ and $\varM$. They are summarised in the diagram of Fig.~\ref{fig-relationbetweenthederivations}.

\begin{figure}
\centering $\xymatrix@R=15pt@C=15pt{
    &
    & 
    {\algzero}
    \ar[d]  
    & 
    {\algzero}  
    \ar[d]   
    &    
    \\ 
    & 
    {\algzero}
    \ar[d]    
    \ar[r]  
    & 
    {\caZ_{\der}(\algA)}
    \ar[d]   
    \ar[r]     
    & 
    {\Gamma(V\varP)}
    \ar[d]
    \ar[r]    
    & 
    {\algzero}  
    \\ 
    {\algzero} 
    \ar[r]
    & 
    {\Int(\algA)}
    \ar[d]    
    \ar[r]
    & 
    {\caN_{\der}(\algA)} 
    \ar[d]^{\tau}
    \ar[r]^{\rho_\varP}   
    & 
    {\Gamma_{\varM}(\varP)}
    \ar[d]^{\pi_\ast}
    \ar[r]  
    & 
    {\algzero} 
    \\ 
    {\algzero} 
    \ar[r]
    & 
    {\Int(\algA)}
    \ar[d]   
    \ar[r]
    & 
    {\der(\algA)}
    \ar[d]
    \ar[r]^{\rho}  
    & 
    {\Gamma(\varM)}
    \ar[d]
    \ar[r]
    & 
    {\algzero}
    \\ 
    & 
    {\algzero}
    & 
    {\algzero}
    & 
    {\algzero}
    &  
    }$
\caption{Some relations between the derivations of $\algB$ and $\algA$ and some vector fields on $\varP$ and $\varM$.}
\label{fig-relationbetweenthederivations}
\end{figure}

In this diagram, one has the following short exact sequences of Lie algebras and $C^\infty(\varM)$-modules:
\begin{itemize}
\item $\xymatrix@1@C=15pt{{\algzero} \ar[r] & {\Int(\algA)} \ar[r] & {\der(\algA)} \ar[r]^-{\rho} & {\Gamma(\varM)} \ar[r] & {\algzero}}$

This is the short exact sequence which relates vector fields on $\varM$, derivations on $\algA$ and inner derivations on $\algA$ given in Proposition~\ref{prop-thederivationsodalgA}.

\item $\xymatrix@1@C=15pt{{\algzero} \ar[r] & {\caZ_{\der}(\algA)} \ar[r] & {\caN_{\der}(\algA)} \ar[r]^-{\tau} & {\der(\algA)} \ar[r] & {\algzero}}$ 

$\caN_{\der}(\algA) \subset \der(\algB)$ is the subset of derivations on $\algB$ which preserve $\algA \subset \algB$.\\
$\caZ_{\der}(\algA) \subset \der(\algB)$ is the subset of derivations on $\algB$ which vanish on $\algA$. This is a Lie ideal in $\caN_{\der}(\algA)$, and $\tau$ is the quotient map.\\
The Lie algebra $\caZ_{\der}(\algA)$ is generated as a $C^\infty(\varP)$-module by the elements $\xi^v + \adrep_\xi$ for any $\xi \in \ksu(n)$.

\item $\xymatrix@1@C=15pt{{\algzero} \ar[r] & {\Gamma(V\varP)} \ar[r] & {\Gamma_{\varM}(\varP)} \ar[r]^-{\pi_\ast} & {\Gamma(\varM)} \ar[r] & {\algzero}}$ 

These are pure geometrical objects:\\
$\Gamma(V\varP)$ is the Lie algebra of vertical vector fields on $\varP$.\\
$\Gamma_{\varM}(\varP)= \{ \caX \in \Gamma(\varP) / \pi_\ast \caX(p) = \pi_\ast \caX(p') \ \forall p , p' \in \varP \text{ s.t. }  \pi(p)=\pi(p')     \}$ is the Lie algebra of vector fields on $\varP$ which can be mapped to vector fields on $\varM$ using the tangent maps $\pi_\ast : T_p\varP \rightarrow T_{\pi(p)}\varM$.

\item $\xymatrix@1@C=15pt{{\algzero} \ar[r] & {\Int(\algA)} \ar[r] & {\caN_{\der}(\algA)} \ar[r]^-{\rho_\varP} & {\Gamma_{\varM}(\varP)} \ar[r] & {\algzero}}$

Here, the elements in $\Int(\algA)$ are identified to the $\adrep_\gamma$ for $\gamma \in \algA_0 \subset \algB$. $\Int(\algA)$ is then a Lie subalgebra of $\caN_{\der}(\algA)$.\\
$\rho_\varP$ is the restriction to $\caN_{\der}(\algA)$ of the projection on the first term in the splitting $\der(\algB) = \Gamma(\varP) \oplus [ C^\infty(\varP) \otimes \der(M_n) ]$.

\end{itemize}

An ordinary connection $\omega \in \Omega^1(\varP)\otimes\ksu(n)$ splits these short exact sequences. Let us look more closely at the central square of the diagram of Fig.~\ref{fig-relationbetweenthederivations}. One can define splittings as follows:
\begin{equation*}
    \xymatrix@R=0pc@C=0pc@M=2pt{ 
      \caN_{\der}(\algA) 
      \ar@{->>}[rrrrrrr]^{\rho_\varP} 
      \ar@{->>}[dddddd]_{\tau} 
      &&&& 
      \rule{60pt}{0pt} 
      &&& 
      \Gamma_\varM(\varP)
      \ar@{->>}[dddddd]^{\pi_\ast} \\
      &&&
      \makebox[75pt][r]{$(\pi_\ast\caX)^h + \omega(\caX)^v + \adrep_{\omega(\caX)}$}
      & & 
      \caX
      \ar@{|->}[ll]
      &&\\
      &
      \makebox[15pt][l]{$\rho(\kX)^h - \adrep_{\alpha(\kX)^\algB}$}
      &&&&& 
      X^h
      & \\
      &&&&
      \rule{0pt}{30pt}
      &&&\\
      &
      \kX
      \ar@{|->}[uu]
      &&&&& 
      X
      \ar@{|->}[uu]
      & \\
      &&
      \nabla_X
      &&& 
      X
      \ar@{|->}[lll]
      &&\\
      \der(\algA) \ar@{->>}[rrrrrrr]_{\rho} 
      &&&&&&& 
      \Gamma(\varM)
      }
\end{equation*}
where:
\begin{itemize}
\item $\Gamma(\varM) \rightarrow \der(\algA)$, $X \mapsto \nabla_X$:

This is the splitting mentioned in Proposition~\ref{prop-Thenoncommutative1formalpha}, which lifts vector fields on $\varM$ into derivations on $\algA$.

\item $\Gamma(\varM) \rightarrow \Gamma_{\varM}(\varP)$, $X \mapsto X^h$:

This splitting lifts vector fields on $\varM$ into horizontal vector fields $X^h$ on $\varP$ through the ordinary geometrical procedure. Using its equivariance, one can easily verify that the vector field $X^h$ is indeed a $\pi_\ast$-projectable vector field. In fact, for any $\caX \in \Gamma_{\varM}(\varP)$, one has $\caX = (\pi_\ast \caX)^h + \caX^v$, where $\caX^v$ is the vertical projection of $\caX$, explicitly given by the formula $\caX^v = \omega(\caX)^v$.

\item $\der(\algA) \rightarrow \caN_{\der}(\algA)$, $\kX \mapsto \rho(\kX)^h - \adrep_{\alpha(\kX)^\algB}$:

This lifts derivations on $\algA$ into derivations on $\algB$. Here, $\alpha(\kX)^\algB$ is the basic element in $\algB$ associated to $\alpha(\kX) \in \algA$ and $\rho(\kX)^h \in \Gamma(\varP)$ is the horizontal lift of the vector field $\rho(\kX)$. By construction, one has $\adrep_{\alpha(\kX)^\algB} \in \caN_{\der}(\algA)$. On the other hand, one verifies that for any $X \in \Gamma(\varM)$ and any $\xi \in \ksu(n)$, $[\xi^v + \adrep_\xi, X^h] = 0$ (as an element in $\der(\algB)$), which shows that $X^h \in \caN_{\der}(\algA)$. The relation $\tau(\rho(\kX)^h - \adrep_{\alpha(\kX)^\algB}) = \kX$ relies on the two following facts: in the identification of $a \in \algA$ as an equivariant map $a^\algB : \varP \rightarrow M_n$, one has the identification of $\nabla_X a$ with $X^h \cdotaction a^\algB$, and one has the decomposition $\kX = \nabla_{\rho(\kX)} - \adrep_{\alpha(\kX)}$.

\item $\Gamma_{\varM}(\varP) \rightarrow \caN_{\der}(\algA)$, $\caX \mapsto (\pi_\ast\caX)^h + \omega(\caX)^v + \adrep_{\omega(\caX)}$:

Here, we lift $\pi_\ast$-projectable vector fields $\caX$ on $\varP$ into derivations on $\algB$. Notice that for any $\caX \in \Gamma_{\varM}(\varP)$, one has the decomposition $\caX = (\pi_\ast\caX)^h + \omega(\caX)^v$. The inner derivation $\adrep_{\omega(\caX)}$ is there in order that $\omega(\caX)^v + \adrep_{\omega(\caX)} \in \caN_{\der}(\algA)$ (we know from the previous result that $(\pi_\ast\caX)^h \in \caN_{\der}(\algA)$). In fact, one has the more interesting result that
\begin{equation*}
\omega(\caX)^v + \adrep_{\omega(\caX)} \in \caZ_{\der}(\algA)
\end{equation*}

\end{itemize}

In order to better understand the two liftings ending in $\caN_{\der}(\algA)$, it is useful to characterize derivations in $\caN_{\der}(\algA)$. Such a derivation can be decomposed, as an element in $\der(\algB)$, as $\widehat{\kX} = \caX + \adrep_b$, with $\caX \in \Gamma(\varP)$ and $b \in \algB_0 = C^\infty(\varP) \otimes \ksl_n$. Using the fact that $\rho_\varP$ is just the restriction of $\widehat{\kX}$ to $C^\infty(\varP)$, one has $\rho_\varP(\widehat{\kX}) = \caX \in \Gamma_{\varM}(\varP)$. The condition $\widehat{\kX} \in  \caN_{\der}(\algA)$ implies that $[\xi^v + \adrep_\xi, \widehat{\kX}] \in \caZ_{\der}(\algA)$ for any $\xi \in \ksu(n)$. Using the structure of $\caZ_{\der}(\algA)$, one can write $[\xi^v + \adrep_\xi, \widehat{\kX}] = f^i (\eta_i^v + \adrep_{\eta_i})$ for some $f^i \in C^\infty(\varP)$ and $\eta_i \in \ksu(n)$, which can be decomposed into two parts: $[\xi^v, \caX] = f^i \eta_i^v$ and $\xi^v \cdotaction b + [\xi, b] = f^i \eta_i$. 

Denote by $L^\equ_\xi = L_{\xi^v} + \adrep_\xi$ the Lie derivative associated to the Cartan operation of $\kg_\equ$ on $(\Omega^\grast_\der(\algB), \hd)$. The second relation is then $L^\equ_\xi b = f^i \eta_i$. Applying now the connection $1$-form $\omega$ on the first relation, one gets $\omega([\xi^v, \caX]) = f^i \eta_i$, which can be written, using the equivariance of $\omega$: $L^\equ_\xi \omega(\caX) = f^i \eta_i$. The difference $a(\widehat{\kX}) = \omega(\caX) - b$ is then $L^\equ$-invariant, which means that $a(\widehat{\kX}) \in \algA$. With $\kX = \tau(\widehat{\kX})$, this is exactly the element $\alpha(\kX) \in \algA$ identified as an element in $\algB$, where $\alpha$ is the noncommutative $1$-form associated to the connection $\omega$.

Using these constructions, one has the following decomposition of any $\widehat{\kX} \in \caN_{\der}(\algA)$:
\begin{equation*}
\widehat{\kX} = \caX + \adrep_b = (\pi_\ast\caX)^h + \underbrace{\omega(\caX)^v + \adrep_{\omega(\caX)}}_{\in \caZ_{\der}(\algA)} - \underbrace{\adrep_{a(\widehat{\kX})}}_{ \in \Int(\algA)}
\end{equation*}



\section{Cohomology and characteristic classes}

In ordinary differential geometry, it is possible to relate the cohomology of a fiber bundle to the cohomology of its base manifold using a spectral sequence based on a Čech-de~Rham bicomplex constructed using differential forms. We will show that such a construction can be performed with the space of noncommutative differential forms.

Using the noncommutative geometry structures described above, it is also possible to recover the Chern characteristic classes of the vector bundle $\varE$. The construction we present in the following is purely algebraic, and relies on an adaptation of some work by Lecomte about characteristic classes associated to splitting of short exact sequence of Lie algebras.

\subsection{\texorpdfstring{The cohomology of $\Omega^\grast_\der(\algA)$}{The cohomology of Omegagrastder(algA)}}

Let us recall the Leray theorem in ordinary differential geometry.

\begin{theorem}[Leray]
For any fiber bundle $\fibre[\pi]{\varF}{\varE}{\varM}$, there exists a spectral sequence $\{ \modE_r\}$ converging to the cohomology of the total space $H^\grast_{\text{dR}}(\varE)$ with  
\begin{equation*}
\modE_2^{p,q} = H^p( \kU ; \faiscH^q) 
\end{equation*}
where $\faiscH^q(U) = H_{\text{dR}}^q( \pi^{-1} U)$ is a locally constant presheaf on the good covering $\kU$ of $\varM$.

If $\varM$ is simply connected and $H^q_{\text{dR}}(\varF)$ is finite dimensional, then
\begin{equation*}
\modE_2^{p,q} = H_{\text{dR}}^p(\varM) \otimes H^q_{\text{dR}}(\varF)
\end{equation*}
\end{theorem} 

One of the proofs of this theorem relies on the construction of a Čech-de~Rham bicomplex as illustrated in the diagram of Fig.~\ref{fig-CechdeRhambicomplex} (see \cite{BottTu:95} for instance):
\begin{equation*}
\modK^{p,q} = \prod_{\alpha_0 < \dots < \alpha_p} \Omega^q( \varE_{U_{\alpha_0 \dots \alpha_p}}) = \prod_{\alpha_0 < \dots < \alpha_p} \Omega^q( \pi^{-1} U_{\alpha_0 \dots \alpha_p})
\end{equation*}
with $U_{\alpha_0 \dots \alpha_p} = U_{\alpha_0} \cap \cdots \cap U_{\alpha_p}$ for $U_{\alpha_i} \in \kU$, where $\kU$ is a good cover of $\varM$, $\dd : \modK^{p,q} \rightarrow \modK^{p,q+1}$ is the ordinary de~Rham differential on the spaces $\Omega^\grast( \varE_{U_{\alpha_0 \dots \alpha_p}})$, and $\delta : \modK^{p,q} \rightarrow \modK^{p+1,q}$ is the Čech differential
\begin{equation*}
(\delta \omega_p)_{\alpha_0 \dots \alpha_{p+1}} = \sum_{i=0}^{p+1} (-1)^i {\omega_{\alpha_0 \dots \widehat{\alpha_i} \dots \alpha_{p+1}}}_{| U_{\alpha_0 \dots \alpha_{p+1}}}
\end{equation*}

\begin{figure}
\centering ${\fontsize{10pt}{10.7pt}\selectfont%
\xymatrix@R=12pt@C=12pt{
 & \vdots & \vdots & \vdots & \dots & \vdots & \dots \\
\algzero \ar[r] & {\textstyle \Omega^q(\varM)} \ar[r]^-{\delta} \ar[u]^-{\dd} & {\textstyle\prod \Omega^q(U_{\alpha_0})} \ar[r]^-{\delta} \ar[u]^-{\dd} & {\textstyle \prod \Omega^q( U_{\alpha_0 \alpha_1})} \ar[r]^-{\delta} \ar[u]^-{\dd} & {\dots} \ar[r]^-{\delta} & {\textstyle \prod \Omega^q( U_{\alpha_0 \dots \alpha_p}) } \ar[r]^-{\delta} \ar[u]^-{\dd} & {\dots} \\
{\vdots} & {\vdots} \ar[u]^-{\dd} & {\vdots} \ar[u]^-{\dd} & {\vdots} \ar[u]^-{\dd} & {\dots} & {\vdots} \ar[u]^-{\dd} & {\dots} \\
\algzero \ar[r] & {\textstyle \Omega^2(\varM)} \ar[u]^-{\dd} \ar[r]^-{\delta} & {\textstyle \prod \Omega^2(U_{\alpha_0})} \ar[u]^-{\dd} \ar[r]^-{\delta} & {\textstyle \prod \Omega^2( U_{\alpha_0 \alpha_1})} \ar[u]^-{\dd} \ar[r]^-{\delta} & {\dots} \ar[r]^-{\delta} & {\textstyle \prod \Omega^2( U_{\alpha_0 \dots \alpha_p}) } \ar[u]^-{\dd} \ar[r]^-{\delta} & {\dots} \\
\algzero \ar[r] & {\textstyle \Omega^1(\varM)} \ar[u]^-{\dd} \ar[r]^-{\delta} & {\textstyle \prod \Omega^1(U_{\alpha_0})} \ar[u]^-{\dd} \ar[r]^-{\delta} & {\textstyle \prod \Omega^1( U_{\alpha_0 \alpha_1})} \ar[u]^-{\dd} \ar[r]^-{\delta} & {\dots} \ar[r]^-{\delta} & {\textstyle \prod \Omega^1( U_{\alpha_0 \dots \alpha_p}) } \ar[u]^-{\dd} \ar[r]^-{\delta} & {\dots} \\
\algzero \ar[r] & {\textstyle \Omega^0(\varM)} \ar[u]^-{\dd} \ar[r]^-{\delta} & {\textstyle \prod \Omega^0(U_{\alpha_0})} \ar[u]^-{\dd} \ar[r]^-{\delta} & {\textstyle \prod \Omega^0( U_{\alpha_0 \alpha_1})} \ar[u]^-{\dd} \ar[r]^-{\delta} & {\dots} \ar[r]^-{\delta} & {\textstyle \prod \Omega^0( U_{\alpha_0 \dots \alpha_p}) } \ar[u]^-{\dd} \ar[r]^-{\delta} & {\dots} \\
 & & {\textstyle C^0(\kU; \gR)} \ar[u]^{i} \ar[r]^-{\delta} & {\textstyle C^1(\kU; \gR)} \ar[u]^-{i} \ar[r]^-{\delta} & {\dots} \ar[r]^-{\delta} & {\textstyle C^p(\kU; \gR)} \ar[u]^-{i} \ar[r]^-{\delta} & {\dots} \\
 & & {\algzero} \ar[u] & {\algzero} \ar[u] & {\dots} & {\algzero} \ar[u] }%
}$
\caption{The ordinary Čech-de~Rham bicomplex associated to a fiber bundle $\fibre[\pi]{\varF}{\varE}{\varM}$}
\label{fig-CechdeRhambicomplex}
\end{figure}

\bigskip
One can introduce a noncommutative Čech-de~Rham bicomplex for $\algA$. In order to do that, denote by $\algA(U) \simeq C^\infty(U) \otimes M_n$ the sections of $\End(\varE)$ restricted over a local trivialisation $U \subset \varM$ with $U \in \kU$, where as before $\kU$ is a good cover of $\varM$. Denote by $g_{UV} : U \cap V \rightarrow SU(n)$ the transition functions for $\varE$. 

For any noncommutative $p$-form $\omega = a_0 \hd a_1 \cdots \hd a_p \in \Omega^p_\der(\algA(U))$ and any differential function $g : U \rightarrow SU(n)$, define the action of $g$ on $\omega$ by $\omega^g = (g^{-1} a_0 g) \hd (g^{-1} a_1 g) \cdots \hd (g^{-1} a_p g)$.

\begin{lemma}[The presheaf $\Omega^\grast_\der(\algA(U))$]
For any $V \subset U$, the maps $i^V_U : \Omega^\grast_\der(\algA(U)) \rightarrow \Omega^\grast_\der(\algA(V))$ given by $\omega \mapsto (\omega_{|V})^{g_{UV}}$ (restriction to $V$ and action of $g_{UV}$) give to $U \mapsto \Omega^\grast_\der(\algA(U))$ a structure of presheaf on $\varM$, which we denote by $\faiscF$.
\end{lemma} 

Using this presheaf, one can introduce the bicomplex 
\begin{equation*}
\modC^{p,q}(\kU; \faiscF) = \prod_{\alpha_0 < \cdots < \alpha_p} \Omega^q_\der(\algA(U_{\alpha_0 \dots \alpha_p}))
\end{equation*}
where by convention the trivialisation over $U_{\alpha_0 \dots \alpha_p}$ is the one over $U_{\alpha_p}$. Let $\hd : \modC^{p,q} \rightarrow \modC^{p,q+1}$ be the noncommutative differential, and define $\delta : \modC^{p,q}(\kU; \faiscF) \rightarrow \modC^{p+1,q}(\kU; \faiscF)$ by (here $g_{\alpha \beta} = g_{U_\alpha U_\beta}$)
\begin{equation*}
(\delta \omega)_{\alpha_0 \dots \alpha_{p+1}} = \sum_{i=0}^p (-1)^i (\omega_{\alpha_0 \dots \widehat{\alpha_i} \dots \alpha_{p+1}})_{|U_{\alpha_0 \dots \alpha_{p+1}}}
 + (-1)^{p+1} (\omega_{\alpha_0 \dots \alpha_{p}})_{|U_{\alpha_0 \dots \alpha_{p+1}}}^{g_{\alpha_p \alpha_{p+1}}}
\end{equation*}
Notice that in the last term, the action of $g_{\alpha_p \alpha_{p+1}}$ performs the change of trivialisation from the one above $U_{\alpha_p}$ to the one above $U_{\alpha_{p+1}}$.

Denote by $\modC^{-1,q}(\kU; \faiscF) = \Omega^q_\der(\algA)$ and define $\delta : \modC^{-1,q}(\kU; \faiscF) \rightarrow \modC^{0,q}(\kU; \faiscF)$ as the restrictions to the trivialisations of the good cover.

One has the following results about the cohomology of $\Omega^\grast_\der(\algA)$:

\begin{proposition}[Noncommutative Leray theorem]
The cohomology of the total complex of the bicomplex $(\modC^{\grast,\grast}(\kU; \faiscF), \hd, \delta)$ is the cohomology of $\Omega^\grast_\der(\algA)$.

The spectral sequence $\{ \modE_r \}$ associated to the filtration 
\begin{equation*}
F^p \modC(\kU; \faiscF) = \textstyle\bigoplus_{s \geq p} \bigoplus_{q \geq 0} \modC^{s,q}(\kU; \faiscF)
\end{equation*}
 converges to the cohomology of $\Omega^\grast_\der(\algA)$ and satisfies 
\begin{equation*}
\modE_2 = H_{\text{dR}}^\grast(\varM) \otimes \caI(\exter^\grast\ksl_n^\ast)
\end{equation*}
\end{proposition}

Recall that the structure of $\caI(\exter^\grast\ksl_n^\ast)$ is known. One can find the proof of this result in \cite{Mass:15}.

\subsection{Characteristic classes and short exact sequences of Lie algebras}

Let us now show that the splitting \eqref{eq-splittingofshortexactsequenceofderivations} of the short exact sequence of derivations contains all the informations needed to recover the Chern characteristic classes of the fiber bundle $\varE$. In order to do that, one has first to introduce a construction by Lecomte (see \cite{Leco:85}).

Let $\xymatrix@1@C=15pt{{\algzero} \ar[r] & {\ki} \ar[r] & {\kg} \ar[r]^-{\pi} & {\kh} \ar[r] & {\algzero}}$ be a short exact sequence of Lie algebras, and let $\varphi : \kh \rightarrow \kg$ be a morphism which splits it as vector spaces. Define $R_\varphi = \dd_{\kh}\varphi + \frac{1}{2}[ \varphi, \varphi] : \exter^2 \kh^\ast \otimes \kg$ with $\dd_{\kh}$ the differential on $\exter^\grast \kh^\ast \otimes \kg$ for the trivial representation of $\kh$ on $\kg$. For any $x,y \in \kh$, the quantity $R_\varphi(x,y) = - \varphi([x,y]) + [\varphi(x), \varphi(y)]$ is exactly the obstruction on $\varphi$ to be a Lie algebra morphism, \textsl{i.e.} a splitting of Lie algebras.

It is worth to note that $R_\varphi$ looks like a curvature, and indeed, the following construction treats it as if it were a curvature. One can show that $R_\varphi$ belongs to $\exter^2 \kh^\ast \otimes \ki$ and that it satisfies a Bianchi identity $d_\kh R_\varphi + [\varphi, R_\varphi] = 0$.

Now, let $\evV$ be a vector space and $\rho$ a representation of $\kh$ on $\evV$. Denote by $\symes^q_\rho(\ki, \evV)$ the space of linear symmetric maps $\otimes^{q} \ki \rightarrow \evV$ which intertwine the adjoint representation $\adrep^{\otimes^q}$ of $\kg$ on $\otimes^{q} \ki$ and the representation $\rho \circ \pi$ of $\kg$ on $\evV$. Let $\epsilon$ be the antisymmetrisation map $\otimes^\grast \kh^\ast \rightarrow \exter^\grast \kh^\ast$.

One has the following result, shown in \cite{Leco:85}:
\begin{proposition}[Characteristic classes of a short exact sequence of Lie algebras]
For any $\alpha \in \symes^q_\rho(\ki, \evV)$, let $\alpha_\varphi = \epsilon \circ \alpha (R_\varphi \otimes \cdots \otimes R_\varphi) \in \exter^{2q} \kh^\ast \otimes \evV$. Then one has $\dd \alpha_\varphi = 0$ where $\dd$ is the differential of the complex $\exter^\grast \kh^\ast \otimes \evV$.

The cohomology class of $\alpha_\varphi$ in $H^{2q}(\kh;\evV)$ does not depends on the choice of $\varphi$.

If the short exact sequence is split exact as a Lie algebra short exact sequence then this cohomology class is zero. 
\end{proposition} 

Let us adapt this construction to the short exact sequence 
\begin{equation*}
\xymatrix@1@R=0pt@C=15pt{ {\algzero} \ar[r] & {\Int(\algA)} \ar[r] & {\der(\algA)} \ar[r]^-{\rho} & {\Gamma(\varM)} \ar[r] & {\algzero}}
\end{equation*}
It is possible to generalise the previous construction in order to take into account the extra structures of $\caZ(\algA)$-modules.

We identify $\Int(\algA)$ with $\algA_0$. The adjoint representation of $\der(\algA)$ on $\Int(\algA)$ is given by $\adrep_\kX (\adrep_a) = [\kX, \adrep_a] = \adrep_{\kX(a)}$ so that it is $(\kX, a) \mapsto \kX(a)$ on $\algA_0$.

The vector space (and $\caZ(\algA)$-module) we consider is $\caZ(\algA)$ itself, on which the representation $\rho$ is $(\kX, f) \mapsto \rho(\kX)\cdotaction f$.

Let $\symes^q_{\caZ(\algA)}(\algA_0, \caZ(\algA))$ be the space of $\caZ(\algA)$-linear symmetric maps $\otimes_{\caZ(\algA)}^{q} \algA_0 \rightarrow \caZ(\algA)$ which intertwine the adjoint representation $\adrep^{\otimes^q}$ of $\der(\algA)$ on $\otimes_{\caZ(\algA)}^{q} \Int(\algA) = \otimes_{\caZ(\algA)}^{q} \algA_0$ and the representation $\rho$ of $\der(\algA)$ on $\caZ(\algA)$.

Notice that, thanks to the $\caZ(\algA)$-linearity, maps in $\symes^q_{\caZ(\algA)}(\algA_0, \caZ(\algA))$ are local on $\varM$, so that one can look at them in local trivialisations of $\varE$. In such a trivialisation over an open set $U$, the intertwining relations can be written, with the usual notation $\kX^\loc = X + \adrep_\gamma$:
\begin{align*}
\sum_{i=1}^q \phi(a_1 \otimes \cdots \otimes X \cdotaction a_i \otimes \cdots \otimes a_q) &= X \cdotaction \phi( a_1 \otimes \cdots \otimes a_q)\\
\sum_{i=1}^q \phi(a_1 \otimes \cdots \otimes [\gamma, a_i] \otimes \cdots \otimes a_q) &= 0
\end{align*}
for any $a_i : U \rightarrow \ksl_n$.

\begin{proposition}[Characteristic classes of $\varE$]
The space $\symes^q_{\caZ(\algA)}(\algA_0, \caZ(\algA))$ is well defined, which means that the $\caZ(\algA)$-linearity and the intertwining condition are compatible, and one has
\begin{equation*}
\symes^q_{\caZ(\algA)}(\algA_0, \caZ(\algA)) = \caP^q_I(\ksl_n)
\end{equation*}
the space of invariant polynomials on the Lie algebra $\ksl_n$.

The differential complex in which the characteristic classes for the splitting are defined is
\begin{equation*}
\Hom_{\caZ(\algA)}(\exter_{\caZ(\algA)}^\grast \Gamma(\varM), \caZ(\algA))
\end{equation*}
which is the de~Rham complex of differential forms on $\varM$.

The characteristic classes one computes in this way are precisely the ordinary Chern characteristic classes of the vector bundle $\varE$ (or of the principal fiber bundle $\varP$).
\end{proposition} 

The last statement relies on the fact that any ordinary connection $\nabla^\varE$ on $\varE$ gives rise to a splitting of the short exact sequence whose curvature is exactly the obstruction to be a morphism of Lie algebras. The construction based on $\symes^q_{\caZ(\algA)}(\algA_0, \caZ(\algA)) = \caP^q_I(\ksl_n)$ is then the ordinary Chern-Weil morphism.


\section{Invariant noncommutative connections}

Many works have been done in the theory of ordinary connections which are symmetric with respect to the action of a Lie group. This leads to understand some ansatz used to get exact solutions of Yang-Mills theories, and recover or introduce some Yang-Mills models coupled with scalar fields through these symmetric reductions. 

In this section, we generalize these considerations to noncommutative connections on the algebra $\algA$, and show that the geometrico-algebraic structures introduced so far are very natural in the theory of symmetric reductions.

This exposé is based on \cite{Mass:25}, and we refer to this paper for more details and references.

\subsection{Action of a Lie group on a principal fiber bundle}

Let us recall some general constructions that were introduced in the theory of symmetric reduction of connections.

Let $\fibre[\pi]{G}{\varP}{\varM}$ be a $G$-principal fiber bundle, and let $H$ be a Lie group acting on the left on $\varP$, such that the action commutes with the right action of $G$. 

Then, the action of $H$ on $\varP$ induces a left action of $H$ on $\varM$. In the following, we assume that this action is simple, which means that $\varM$ admits the fiber bundle structure $\fibre{H/H_0}{\varM}{\varM/H}$ where $H_0$ is an isotropy subgroup: $H_0 = H_{x_0} = \{ h \in H \ / \ h\cdotaction x_0 = x_0\}$. In particular, all the isotropy subgroups are isomorphics to one of them. We fix $H_0$ as such an isotropy subgroup.

Then we introduce the following spaces:

\begin{itemize}
\item $\varN = \{ x \in \varM \ / \ H_x = H_0 \}$ is the space of points in $\varM$ whose isotropy subgroup is exactly $H_0$.

\item $\caN_H(H_0) = \{ h \in H \ / \ h H_0 = H_0  h \}$ is the normalizer of $H_0$ in $H$.

\item $H_0$ is a normal subgroup of $\caN_H(H_0)$, and one has the principal fiber bundle
\begin{equation*}
\fibre{\caN_H(H_0)/H_0}{\varN}{\varM/H}
\end{equation*}
The fiber bundle $\fibre{H/H_0}{\varM}{\varM/H}$ is associated to this bundle for the natural action of $\caN_H(H_0)/H_0$ on $H/H_0$ by (right) multiplication.
\end{itemize}

Define $S = H \times G$. This group acts on the right on $\varP$ by the following relation: $(h,g)\cdotaction p = h^{-1}\cdotaction p \cdotaction g$. For any $p \in \varP$, let $\lambda_p : H_{\pi(p)} \rightarrow G$ be defined such that $h \cdotaction p = p \cdotaction \lambda_p(h)$. Then one can show the followings:
\begin{itemize}
\item $S_p = \{ (h, \lambda_p(h)) \ / \ h \in H_{\pi(p)} \}$ is the isotropy subgroup of $p \in \varP$ for the action of $S$. This implies that the action of $S$ on $\varP$ is simple.
\item Fix an isotropy subgroup $S_0$ and let $\varQ = \{ p \in \varP \ / \ S_p = S_0 \}$. Then
\begin{equation*}
\fibre{S/S_0}{\varP}{\varM/H} \text{ is associated to } 
\fibre{\caN_S(S_0)/S_0}{\varQ}{\varM/H}
\end{equation*}
as for the (simple) action of $H$ on $\varM$.
\end{itemize}

\begin{proposition}[Some properties of $\varQ$ and $\lambda$]
The map $\lambda_p : H_{\pi(p)} \rightarrow G$ such that $h \cdotaction p = p \cdotaction \lambda_p(h)$ satisfies
\begin{equation*}
\lambda_{p \cdotaction g}(h) = g^{-1} \lambda_p(h) g
\end{equation*}

For any $q \in \varQ$, $\lambda_q$ depends only on $\pi(q) \in \varM$: $\lambda_q(h) = \lambda_{q\cdotaction g}(h)$. For a fixed $x_0$ in $\varM$ whose isotropy group is $H_0$, we denote this map restricted to $\varQ$ by $\lambda : H_0 \rightarrow G$.

The projection $\pi : \varP \rightarrow \varM$ induces the fiber bundle structure
\begin{equation*}
\fibre[\pi_\varQ]{\caZ_G(\lambda(H_0))}{\varQ}{\pi(\varQ)}
\end{equation*}
with $\caZ_G(\lambda(H_0)) = \{ g \in G \ / \ g \lambda(h_0) = \lambda(h_0) g, \forall h_0 \in H_0 \}$, the centralizer of $\lambda(H_0)$ in $G$, and $\pi(\varQ) \subset \varN$.
\end{proposition} 

\begin{figure}\small
\centering $\xymatrix@R=7pt@C=0pt@M=6pt{%
\caZ_G(\lambda(H_0)) \ar@{^{(}->}[rrr] \ar@{^{(}->}[ddrr] \ar@{=}[ddd] 
& & & 
\caN_S(S_0)/S_0 \ar@{>>}[rrr] \ar@{^{(}->}[ddrr] \ar'[dd][ddd] 
& & & 
(\caN_S(S_0)/S_0) / \caZ_G(\lambda(H_0)) \ar@{^{(}->}[dr] \ar'[dd][ddd] 
& & \\
& & & & & & & 
\caN_H(H_0)/H_0 \ar@{^{(}->}[dr] \ar'[d][ddd] 
& \\
& & 
G \ar@{^{(}->}[rrr] \ar@{=}[ddd] 
& & & 
S/S_0 \ar@{>>}[rrr] \ar[ddd] 
& & & 
H/H_0 \ar[ddd] \\
\caZ_G(\lambda(H_0)) \ar'[rr][rrr] \ar@{^{(}->}[ddrr] 
& & & 
\varQ \ar@{>>}'[rr][rrr]^-{\pi_\varQ} \ar@{^{(}->}[ddrr] \ar@{>>}'[dd][ddd] 
& & & 
\pi(\varQ) \ar@{^{(}->}[dr] \ar@{>>}'[dd][ddd] 
& & \\
& & & & & & & 
\varN \ar@{^{(}->}[dr] \ar@{>>}'[d][ddd] 
& \\
& & 
G \ar[rrr] 
& & & 
\varP \ar@{>>}[rrr]^-{\pi} \ar@{>>}[ddd] 
& & & 
\varM \ar@{>>}[ddd] 
\\
& & & 
\varM/H \ar@{=}'[rr][rrr] \ar@{=}[ddrr] 
& & & 
\varM/H \ar@{=}[dr] 
& & \\
& & & & & & & 
\varM/H \ar@{=}[dr] 
& \\
& & & & & 
\varM/H \ar@{=}[rrr] 
& & & 
\varM/H 
\\%
}$
\caption{In this diagram, some arrows represent true applications and other arrows are part of diagrams of fibrations, most of them explicitly given before. Some horizontal arrows correspond to the action of $G$ (or subgroups of $G$) and some vertical arrows correspond to actions of groups related to $H$ and $S$.}
\label{fig-diagramoffibrations}
\end{figure}
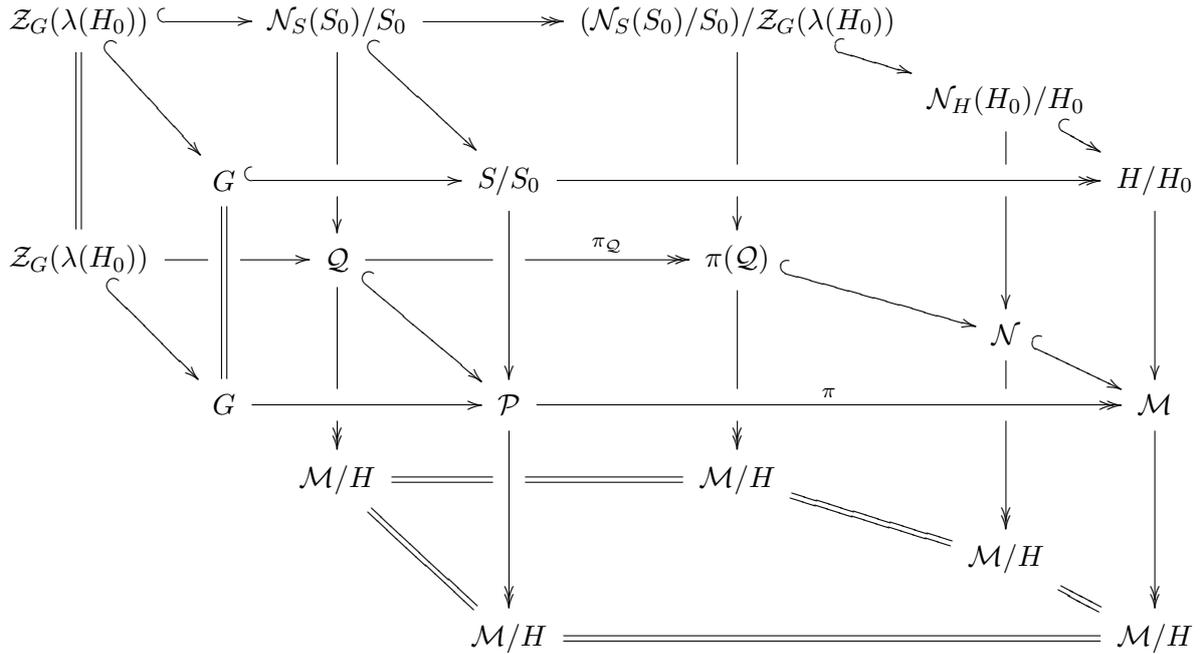

We summarize in Fig.~\ref{fig-diagramoffibrations} all the fibrations one can obtain relating the spaces introduced so far. In the following, we will concentrate more precisely on the diagram of fibrations:
\begin{equation*}
\xymatrix@R=15pt@C=20pt{ 
{\caN_S(S_0)/S_0} \ar@{^{(}->}[d] \ar[r] & {\varQ} \ar@{^{(}->}[d] \ar[r] & {\varM/H} \ar@{=}[d] \\
{S/S_0} \ar[r] & {\varP} \ar[r] & {\varM/H}
}
\end{equation*}

\bigskip
In order to connect this construction to some differential structures, we are now interested in the Lie algebras of the different groups introduced before.

Let us look at the structure of the Lie algebra $\kh$ of the group $H$. One can introduce the following spaces: 
\begin{itemize}
\item $\kh_0$ is the Lie algebra of $H_0$, the once for all fixed isotropy group.
\item $\kk$ is the Lie algebra of the quotient group $\caN_H(H_0)/H_0$.
\item $\kn_0 = \kh_0 \oplus \kk$ is the Lie algebra of $\caN_H(H_0)$, the normalizer of $H_0$ in $H$.
\item $\kl$ is the vector space in the orthogonal decomposition $\kh = \kn_0 \subsetplus \kl$ such that $[\kn_0, \kl] \subset \kl$ (this is called a reductive decomposition of $\kh$ along $\kn_0$).
\end{itemize}

Denote by $\kg$ the Lie algebra of the group $G$. As before, we can introduce the following spaces:
\begin{itemize}
\item $\kz_0$ the Lie algebra of $\caZ_G(\lambda(H_0))$, the centralizer of $\lambda(H_0)$ in $G$.
\item $\km$ the vector space in the orthogonal and reductive decomposition $\kg = \kz_0 \subsetplus \km$ ($[\kz_0, \km] \subset \km$). 
\end{itemize}

The Lie algebra of the group $S = H \times G$ is $\ks = \kh \oplus \kg$, and one has
\begin{itemize}
\item $\ks_0 = \{ (x_0, \lambda_\ast x_0) \ / \ x_0 \in \kh_0 \}$ is the Lie algebra of $S_0$, the fixed isotropy group.
\item $\ks_0 \oplus \kk \oplus \kz_0$ is the Lie algebra of $\caN_S(S_0)$, the normalizer of $S_0$ in $S$.
\item $\kk \oplus \kz_0$ is the Lie algebra of the quotient group $\caN_S(S_0)/S_0$.
\end{itemize}

With these spaces, one has the following result:

\begin{proposition}[Decomposition of $T\varP$]
For any $q \in \varQ$, one has $\kk^\varQ_q \oplus {\kz_0}^\varQ_q \subset T_q \varQ$ and $T_q \varP = T_q \varQ \oplus \kl_q^\varP \oplus \km_q^\varP$.
\end{proposition} 

In this proposition, we use the following compact notation: $\ka_q^\varR$ is the space of tangent vectors over $q$ associated to elements $x \in \ka \subset \kh$ or $\kg$ through the fundamental vector fields on $\varR = \varQ$ or $\varP$ for the action of the corresponding group $H$ or $G$.

\subsection{Invariant noncommutative connections}

It is now possible to mix together the geometrical constructions of the previous subsection and the noncommutative algebraic considerations on the endomorphism algebra associated to a $G$-principal fiber bundle $\varP$ with $G = SL(n)$ or $G = SU(n)$. Let then as before $H$ be a compact connected Lie group acting on $\varP$.

\begin{proposition}[Operations of $\kh$ on $\Omega^\grast_\der(\algB)$ and $\Omega^\grast_\der(\algA)$]
\label{prop-OperationsofkhonOmegagrastderalgBandOmegagrastderalgA}
The operation of $\kh$ on $\Omega^\grast(\varP)$ induced by the action of $H$ on $\varP$ extends to an operation of $\kh$ on $\Omega^\grast_\der(\algB) = \Omega^\grast(\varP) \otimes \Omega^\grast_\der(M_n)$ (using a trivial action on the second factor). This operation commutes with the operations of $\kg_\adrep$ and $\kg_\equ$, and so reduces to the operation of $\kh$ on $\Omega^\grast(\varP)$ and to an operation of $\kh$ on $\Omega^\grast_\der(\algA)$.
\end{proposition} 

\begin{definition}[Invariant noncommutative connection]
The operation of $\kh$ on $\Omega^\grast_\der(\algA)$ obtained in Proposition~\ref{prop-OperationsofkhonOmegagrastderalgBandOmegagrastderalgA} is our definition of the (noncommutative) action of $H$ on the algebra $\algA$.

A noncommutative connection $\widehat{\nabla}$ on the right $\algA$-module $\modM=\algA$ is said to be $\kh$-invariant if, $\forall y \in \kh$, $\forall \kX \in \der(\algA)$ and $\forall a \in \algA$, one has $L_y (\widehat{\nabla}_\kX a) = \widehat{\nabla}_{[y, \kX]} a + \widehat{\nabla}_\kX (L_y a)$, where $L_y$ is the Lie derivative of the operation of $\kh$ on $\Omega^\grast_\der(\algA)$.
\end{definition} 

Using this definition, one obtains the equivalent characterization:
\begin{proposition}[Invariance of the noncommutative $1$-form $\alpha$]
The noncommutative connection $\widehat{\nabla}$ is $\kh$-invariant if and only if its noncommutative $1$-form $\alpha$ is invariant: $L_y \alpha = 0$ for all $y \in \kh$.
\end{proposition} 

This last proposition, combined with the relations between the noncommutative geometries of the algebras $\algA$ and $\algB$, permits one to reduce the problem of finding the $\kh$-invariant noncommutative connections $\widehat{\nabla}$ on the right $\algA$-module $\modM=\algA$ to the following problem: find all the noncommutative $1$-forms written as $\alpha = \omega - \phi \in [\Omega^1(\varP) \otimes M_n] \oplus [C^\infty(\varP) \otimes M_n \otimes \ksl_n^\ast]$ satisfying the four relations
\begin{align*}
(L_{\xi^v} + L_{\adrep_\xi}) \omega &= 0 
&
(L_{\xi^v} + L_{\adrep_\xi}) \phi &= 0 
&
i_{\xi^v} \omega - i_{\adrep_\xi} \phi &= 0
&
L_y( \omega - \phi) &=0
\end{align*}
for all $\xi \in \kg = \ksl_n$ and $y \in \kh$. The three first relations express the basicity of $\alpha$ for the operation of $\kg_\equ$ on $\Omega^\grast_\der(\algB)$, and the last one is the $\kh$-invariance. 

This last relation decomposes into two independent equations $L_y \omega = 0$ and $L_y \phi = 0$ for all $y \in \kh$. This implies in particular that one can restrict the study of $\omega$ and $\phi$ defined over $\varP$ to the submanifold $\varQ \subset \varP$. For all $q \in \varQ$, one has then to characterize the maps
\begin{align*}
\omega_q : T_q \varP= T_q \varQ \oplus \kl_q^\varP \oplus \km_q^\varP &\rightarrow M_n
&
\phi_q : \kg &\rightarrow M_n
\end{align*}

Now, the relation $i_{\xi^v} \omega - i_{\adrep_\xi} \phi = 0$ for all $\xi \in \kg$, says that $\phi_q(\xi)$ is completely determined by $\omega_q(\xi^v_q)$. It is then sufficient to study $\omega_q$.

Let us first consider the $T_q\varQ$ part of $T_q\varP$. Denote by $\mu_q : T_q \varQ \rightarrow M_n$ the restriction of $\omega_q$ to $T_q \varQ$. One has ${\kz_0}^\varQ_q \subset T_q \varQ$, so that $\mu$ and $\phi$ are both defined on $\kz_0 \subset \kg$, where they coincide: $\mu(z^\varQ) = \phi(z)$ for any $z \in \kz_0$. Denote by $\eta_q$ the restriction of $\phi_q$ to $\kz_0$, which is then also the restriction of $\mu_q$ to ${\kz_0}^\varQ_q$. It is possible to write down these relations in a compact way through the following result:
\begin{proposition}[The algebra $\algW$]
Let $\algW = \caZ_{M_n}(\lambda_\ast \kg_0)$ be the centralizer of $\lambda_\ast \kg_0$ in $M_n$. It is an associative algebra and $\kz_0 \subset \der(\algW)$. Let $\Omega^\grast_{\kz_0}(\algW) = \algW \otimes \exter^\grast \kz_0^\ast$ be the restricted derivation-based differential calculus associated to it. 

There is a natural operation of $\kz_0$ on $\Omega^\grast(\varQ) \otimes \Omega^\grast_{\kz_0}(\algW)$, and $\mu - \eta \in \left( \Omega^\grast(\varQ) \otimes \Omega^\grast_{\kz_0}(\algW) \right)^1_{\text{$\kz_0$-basic}}$.
\end{proposition} 

In order to take into account the remaining part of $\mu_q$ in $T_q\varQ$, we introduce the following bigger differential calculus:
\begin{proposition}[The differential calculus $\Omega^\grast_{\kk \oplus \kz_0}(\varM/H; \algW)$]
There are natural operations of $\kk \subset \kh$ and $\kz_0 \subset \kg$ on the differential algebra $\Omega^\grast(\varQ) \otimes \Omega^\grast_{\kz_0}(\algW) \otimes \exter^\grast \kk^\ast$.

Define
\begin{equation*}
\Omega^\grast_{\kk \oplus \kz_0}(\varM/H; \algW) = \left( \Omega^\grast(\varQ) \otimes \Omega^\grast_{\kz_0}(\algW) \otimes \exter^\grast \kk^\ast \right)_{\text{$\kk \oplus \kz_0$-basic}}
\end{equation*}
The algebra $\algC = \Omega^0_{\kk \oplus \kz_0}(\varM/H; \algW) = \left( C^\infty(\varQ) \otimes \algW \right)_{\text{$\kk \oplus \kz_0$-invariants}}$ is the algebra of sections of a $\algW$-fiber bundle associated to the principal fiber bundle
\begin{equation*}
\fibre{\caN_S(S_0)/S_0}{\varQ}{\varM/H}
\end{equation*}
\end{proposition}

Recall that $\kk^\varQ_q \subset T_q \varQ$. For any $k \in \kk$, let us define $\nu_q(k) = \mu_q(k^\varQ_q)$, so that $\nu \in C^\infty(\varQ) \otimes \algW \otimes \exter^1 \kk^\ast$. This element contains the dependence of $\mu_q$ over $\kk^\varQ_q \subset T_q \varQ$, but $\nu$ and ${\mu}_{\mid \kk^\varQ}$ are not equal as elements in $\left( \Omega^\grast(\varQ) \otimes \Omega^\grast_{\kz_0}(\algW) \otimes \exter^\grast \kk^\ast \right)^1$ (they do not have the same tri-graduation).

\begin{proposition}[The $T_q\varQ$ part of $T_q\varP$]
One has $\mu - \eta - \nu \in \Omega^1_{\kk \oplus \kz_0}(\varM/H; \algW)$, and this expression contains all the information about the restriction of $\omega$ to $T\varQ$. 
\end{proposition}

Let us now look at the $\kl_q^\varP \oplus \km_q^\varP$ part of $T_q\varP$.

Recall that $[\kh_0 \oplus \kk, \kl] \subset \kl$ and $[\kz_0, \km] \subset \km$, so that there are natural actions $[\kh_0, \kl \oplus \km] \subset \kl \oplus \km$ and $[\kk \oplus \kz_0, \kl \oplus \km] \subset \kl \oplus \km$. On the other hand, recall that $\ks_0 \oplus \kk \oplus \kz_0$ is the Lie algebra of $\caN_S(S_0)$ and $\kk \oplus \kz_0$ is the Lie algebra of $\caN_S(S_0)/S_0$, with $\ks_0 = \{ (x_0, \lambda_\ast x_0) \ / \ x_0 \in \kh_0 \}$. 

On the restriction of $\omega$ to $\varQ$, the $H$-invariance and the $G$-invariance combine together into a $\caN_S(S_0)$-invariance. One can treat this invariance in two steps: one for $S_0$ and the other one for $\caN_S(S_0)/S_0$.

In order to encode the $S_0$-invariance, let us define the vector space of $S_0$-invariant linear maps $\kl \oplus \km \rightarrow M_n$:
\begin{equation*}
\evF = \{ f : \kl \oplus \km \rightarrow M_n \ / \ f([x_0, v]) - [ \lambda_\ast x_0, f(v)] = 0,\ \forall x_0 \in \kh_0,\ \forall v \in \kl \oplus \km \}
\end{equation*}
on which $\kk \oplus \kz_0$ acts naturally using the Lie derivative $(L_{k+z}f)(v) = - f([k,v]) + [z, f(v)]$ for any $k \in \kk$ and $z \in \kz_0$.

The $\caN_S(S_0)/S_0$-invariance is then encoded into the space $\modM = \left( C^\infty(\varQ) \otimes \evF \right)_{\text{$\kk \oplus \kz_0$-invariants}}$.

\begin{proposition}[The $\kl_q^\varP \oplus \km_q^\varP$ part of $T_q\varP$]
$\modM$ is the space of sections of the $\evF$-fiber bundle associated to the principal fiber bundle 
\begin{equation*}
\fibre{\caN_S(S_0)/S_0}{\varQ}{\varM/H}
\end{equation*}
It is a $\algC$-bimodule.

The restriction of $\omega$ to the subspaces $\kl_q^\varP \oplus \km_q^\varP$ is in $\modM$.
\end{proposition}

Using the two previous decomposition of $\omega$, one get the final identification:
\begin{theorem}[The space of $H$-invariant noncommutative connections]
The space of $H$-invariant noncommutative connections on the endomorphism algebra $\algA$ over the right $\algA$-module $\algA$ is the space $\Omega^1_{\kk \oplus \kz_0}(\varM/H; \algW) \oplus \modM$.
\end{theorem} 

It is important to notice the following facts:
\begin{remark}[Naturality of the spaces]
In this result, all the spaces are constructed from the principal fiber bundle 
\begin{equation*}
\fibre{\caN_S(S_0)/S_0}{\varQ}{\varM/H}
\end{equation*}
with the help of geometrical or algebraic methods which are natural in this noncommutative framework:

\begin{itemize}
\item $\algC = \left( C^\infty(\varQ) \otimes \algW \right)_{\text{$\kk \oplus \kz_0$-invariants}}$ is modeled on the finite dimensional algebra $\algW \subset M_n$. It looks like a ``reduced algebra'' constructed from $\algA$, as $\algA$ itself is a reduced algebra for $\algB$.

\item $\Omega^\grast_{\kk \oplus \kz_0}(\varM/H; \algW)$ is a natural differential calculus over $\algC$.

\item $\modM = \left( C^\infty(\varQ) \otimes \evF \right)_{\text{$\kk \oplus \kz_0$-invariants}}$ is a natural $\algC$-bimodule.

\item The space $SU(\algC)$ acts naturally on the space $\Omega^1_{\kk \oplus \kz_0}(\varM/H; \algW) \oplus \modM$ as restriction of noncommutative gauge transformations.

\item All these spaces are sections of fiber bundles over the base space $\varM/H$.
\end{itemize}
\end{remark} 

We refer to \cite{Mass:25} for examples of such symmetric noncommutative restrictions, in particular the noncommutative generalisation of the well studied situation of spherical $SU(2)$ gauge fields over a flat four dimensional space-time. For purely noncommutative situations, the problem reduces to the study of the decomposition of some representation of $G$ in $M_n$ into irreducible representations.

\bibliography{biblio-articles-perso,biblio-livre,biblio-articles}

\end{document}